\documentclass[prd,twocolumn,amsmath,amssymb,floatfix, preprintnumbers,superscriptaddress,nofootinbib,letterpaper]{revtex4}
\usepackage{color}
\usepackage{multirow}
\usepackage{rotating}
\usepackage{amsmath}
\usepackage{amssymb}
\usepackage{aas_macros}
\newcommand{\dd}{\mathrm{d}}

\newcommand{\Tr}{\mathrm{Tr}~}

\newcommand{\shalfcut}{\mathcal{S}^{\mathrm{cut}}_{1/2}}
\newcommand{\shalf}{\mathcal{S}_{1/2}}
\newcommand{\n}{\hat{\vec{n}}_}

\renewcommand{\vec}[1]{\mathbf{#1}}

\begin{document}
 
\title{The cut-sky cosmic microwave background is not anomalous}

\author{Andrew Pontzen}\email{apontzen@ast.cam.ac.uk}
\affiliation{Institute of Astronomy and Kavli Institute for Cosmology, University of Cambridge, Cambridge CB3 0HA, U.K.}
\author{Hiranya V. Peiris}\email{h.peiris@ucl.ac.uk}
\affiliation{Institute of Astronomy and Kavli Institute for Cosmology, University of Cambridge, Cambridge CB3 0HA, U.K.}
\affiliation{Department of Physics and Astronomy, University College London, London WC1E 6BT, U.K.}

\date{\today}

\newcommand{\hvp}[1]{\textcolor{red}{[{\bf HVP}: #1]}}
\newcommand{\app}[1]{\textcolor{blue}{[{\bf AP}: #1]}}

\begin{abstract}
  The observed angular correlation function of the cosmic microwave
  background has previously been reported to be anomalous,
  particularly when measured in regions of the sky uncontaminated by
  Galactic emission. Recent work by Efstathiou {\it et al.} presents a
  Bayesian comparison of isotropic theories, casting doubt on the
  significance of the purported anomaly. We extend this analysis to
  all anisotropic Gaussian theories with vanishing mean ($\langle
  \delta T \rangle = 0$), using the much wider class of models to
  confirm that the anomaly is not likely to point to new physics.  On
  the other hand if there is any new physics to be gleaned, it results
  from low-$\ell$ alignments which will be better quantified by a
  full-sky statistic.

  We also consider quadratic maximum likelihood power spectrum
  estimators that are constructed assuming isotropy. The underlying
  assumptions are therefore false if the ensemble is anisotropic.
  Nonetheless we demonstrate that, for theories compatible with the
  observed sky, these estimators (while no longer optimal) remain
  statistically superior to pseudo-$C_{\ell}$ power spectrum
  estimators.
\end{abstract}
\pacs{---}

\maketitle

\section{Introduction}

Observations of the cosmic microwave background (CMB) by the {\it
  Wilkinson Microwave Anisotropy Probe} (WMAP; {\it e.g.}
\cite{2003ApJS..148....1B,2010arXiv1001.4744J}) are widely interpreted
as confirming the standard model of cosmology in which inflation
generates a homogeneous and isotropic background and seeds isotropic,
nearly scale-free perturbations.  Yet a variety of tests suggest that,
on large scales, something may be amiss
\cite{Copi:2003kt,Eriksen:2003db,deOliveiraCosta:2003pu,2005PhRvL..95g1301L,2006PhRvD..74b3005D,2007PhRvD..75b3507C,Hansen:2008ym,2008arXiv0808.3767C}. (For
a wide-ranging assessment of such anomalies in the 7-yr WMAP data see
Ref. \cite{Bennett:2010jb}.) The interpretation of these results is
complicated by the {\it a posteriori} nature of anomaly hunting: any
large dataset will contain statistical flukes which, in isolation, can
be made to look unacceptable. This is a particularly pernicious
problem in the context of large-scale cosmology: with only one sky to
observe, frequentist statistics are almost impossible to interpret.

%Bayesian statistics, in which different physical models are compared
%in the light of the entire dataset, avoid some of these
%issues. 
Frequentist results can be made into more concrete Bayesian statements
by considering specific alternative CMB theories or classes of
theories (see {\it e.g.}
Refs. \cite{2009arXiv0911.0150G,2009ApJ...690.1807G,Zheng:2010ty}). But
a single, fixed dataset can still contribute overwhelming evidence in
favour of or against the very same theory, depending on the
alternatives against which we are judging (for an elucidation of this
point, see Ref. \cite{2003prth.book.....J}, Sec. 5.5).  In other words
there is no unique way to ascribe significance to departures from the
standard theory.

This does not imply we should abandon critical evaluations of WMAP and
other data: if we simply accept we have an `unlikely' realization of
our favoured theory, we might miss the opportunity to discover new
physics (or instrumental systematics). Thus frequentist results cannot be
dismissed out-of-hand; but we would advocate their interpretation as
pointers to interesting areas of work, rather than quantifiable
death-knells of existing models or theories.

In the present work, we will consider a long-standing debate about the
nature of the angular correlation function $\mathcal{C}(\theta)$ of
the CMB. The argument is usually phrased in terms of the statistic
$\shalfcut$, which traces the extent to which temperature fluctuations
(outside a Galactic mask) are correlated between points separated by
$60^{\circ}$ or more. For a quantitative definition, see Section
\ref{sec:background-notation}.  A number of recent works have
attempted to assess the significance of the purportedly anomalous
value of $\shalfcut$, reaching essentially contradictory
conclusions. In particular, the frequentist $P$-value
\cite{2008arXiv0808.3767C} suggests the observed sky is highly
anomalous, while a Bayesian analysis of the optimally reconstructed
sky by Efstathiou {\it et al.}  suggests the opposite
\cite{2009arXiv0911.5399E}; see also Ref. \cite{Bennett:2010jb}.
%The analysis in
%Ref. \cite{2009arXiv0911.5399E} considers the full sky to be perfectly
%recoverable from the cut sky, and consequently examines the posterior
%probability of the ensemble-averaged full-sky $\shalf$ given
%the single observed realization. 
However, any Bayesian result pivots crucially on the alternative
models considered; the assumptions in Ref. \cite{2009arXiv0911.5399E}
mean that only isotropic models are considered. This is a significant
omission, since it leaves open the possibility that suboptimal
estimates of $\shalf$ formed from cut sky data can be reframed as
useful measures of anisotropy.

The present work rectifies that omission. The anomaly is analysed from
within harmonic space, and then anisotropic theories which make our
CMB realization more probable are considered.  The $\shalfcut$
anomaly is found to be uninformative in the following two senses:
\begin{enumerate}
\item The trivial maximum likelihood anisotropic Gaussian theory for
  our observed sky\footnote{Namely, that with covariance matrix
    $\mathbf{C}=\vec{a}\vec{a}^{\dagger}$ where $\vec{a}$ is the
    observed sky data vector.} does not lead to substantially better
  likelihoods for the single statistic $\shalfcut$;
\item Theories constructed specifically to maximize the likelihood of
  $\shalfcut$ (ignoring the rest of the information on the sky) also
  yield little gain.
\end{enumerate}
These failures arise from the large variance inherent in using a
statistic, such as $\shalfcut$, which is quartic in the data.
Overall, then, the present work reinforces the view that the
frequentist `unlikeliness' of the observed sky must be regarded as a
statistical fluke.

Some broader results arise from our study. First, we consider the
effect of an anisotropic theory on quadratic maximum likelihood (QML)
estimates of the power spectrum.  The QML estimators are derived
under the (in this context false) assumption of isotropy; despite
this, they typically remain superior to pseudo-$C_{\ell}$ approaches
to power spectrum estimation (Section \ref{sec:background-notation},
with detail in Appendix \ref{sec:relat-betw-qml}). Second, we present
an extremely fast method for finding the maximum angular momentum
direction of a CMB map (Appendix \ref{sec:rapid-calc-l2_m}). Third, we
demonstrate that cut-sky correlation functions can be exactly
reproduced from the pseudo-$C_{\ell}$ power spectrum (Appendix
\ref{sec:estim-ctheta-s_12}). This final result, applicable also for
weighted data, has been reported previously \cite{2004PhRvD..69h3524A}
but ignored by recent work; to our knowledge no proof appears in the
existing literature.

The paper is structured as follows. Section
\ref{sec:background-notation} introduces the necessary background and
notation. In Section \ref{sec:why-mathc-small} we consider, from a
harmonic-space perspective, the origin of the low observed
$S_{1/2}^{\mathrm{cut}}$. Anisotropic, Gaussian theories which
reproduce this result are considered in Section \ref{sec:theories},
and show that even the best conceivable fit to the observed CMB makes
no substantial improvement to the $S_{1/2}^{\mathrm{cut}}$
likelihood. Finally, the work is summarized in Section
\ref{sec:conclusions}. Further details and discussion are contained in
appendices.

\section{Background and notation}\label{sec:background-notation}

In this Section we set out the various definitions needed in our
work. Let us start by defining the observed temperature correlation
function $\mathcal{C}(\theta)$ as
\begin{equation}
\mathcal{C}(\theta) = \overline{T(\hat{\vec{n}}_1) T(\hat{\vec{n}}_2)}\textrm{,} \label{eq:sky-ctheta}
\end{equation}
where the overbar denotes averaging over all observed line-of-sight vector pairs
$\hat{\vec{n}}_1$, $\hat{\vec{n}}_2$ satisfying $\hat{\vec{n}}_1 \cdot
\hat{\vec{n}}_2 = \cos \theta$. 
We further define
\begin{equation}
\mathcal{C}_\ell \equiv \frac{1}{2\ell+1} \sum_m \left|a_{\ell m}\right|^2\textrm{,}\label{eq:sky-cls}
\end{equation}
where the $a_{\ell m}$'s are the spherical harmonic coefficients of
the temperature field on the observed sky. A calligraphic
$\mathcal{C}_\ell$ thus denotes the observed power, distinguished from
the theoretical variances $C_\ell$ which we regard as defined by the
relation
\begin{equation}
C_\ell = \langle \mathcal{C}_\ell \rangle\label{eq:theory-cl-definition}
\end{equation}
for isotropic and anisotropic theories alike. 

There is an exact relationship between $\mathcal{C}(\theta)$ and
$\mathcal{C}_\ell$, namely
\begin{equation}
\mathcal{C}(\theta) = \frac{1}{4\pi} \sum_\ell (2\ell+1) \mathcal{C}_\ell P_\ell(\cos \theta) \textrm{,} \label{eq:c-full-leg}
\end{equation}
where $P_\ell$ are the Legendre polynomials. Throughout this paper
in our numerical calculations, we use a finite sum over $2\le \ell \le 30$;
the lower limit discards any contribution from the monopole and dipole,
while we verified that the upper limit is high enough for our results
to converge.

Result (\ref{eq:c-full-leg}) holds regardless of any theoretical
constraints (such as isotropy). Thus the information in the
correlation function is identical to that in the $\mathcal{C}_\ell$'s
observed on the sky.  If desired, one can define the theoretical
correlation function to be the ensemble average of the sky-observed
correlation function, $C(\theta)=\langle \mathcal{C}(\theta)\rangle$.

The purported anomalies relate to
the apparent lack of correlations on large angular scales, quantified by
\begin{equation}
\shalf = \int_{-1}^{1/2} \mathcal{C}(\theta)^2 \sin \theta\, \dd \theta \textrm{.}\label{eq:shalf-definition}
\end{equation}
This quantity is a measure of the extent to which the temperature from
points separated by $60^{\circ}$ or more is correlated. Rather than
evaluate (\ref{eq:shalf-definition}) directly, it is much faster and
numerically more stable to calculate $\shalf$ from the
quadratic form
\begin{equation}
\shalf = \sum_{\ell\ell'} \mathcal{C}_{\ell} \mathcal{C}_{\ell'} s_{\ell\ell'}\label{eq:shalf-quadratic}\textrm{,} 
\end{equation}
where, as above, $\ell$ and $\ell'$ range from $2$ to $30$ in our
numerical calculations and
\begin{equation}
s_{\ell\ell'} = \int_{-1}^{1/2} P_{\ell}(x) P_{\ell'}(x) \dd x\textrm{,} \label{eq:s-matrix}
\end{equation}
which may be computed using well-known recursion relations
({\it e.g.} Appendix C.2 of Ref. \citep{2001PhRvD..64h3003W}; see also Ref. \cite{2008arXiv0808.3767C}).

If one does not trust information inside a specified mask
(for instance due to suspected Galactic contamination), one may
calculate the
correlation function using only the points outside the mask,
\begin{equation}
\mathcal{C}(\theta)^{\mathrm{cut}} \equiv \left. \overline{T(\hat{\vec{n}}_1) T(\hat{\vec{n}}_2)}\right|_{M(\hat{\vec{n}}_1)=M(\hat{\vec{n}}_2)=1} \textrm{,}\label{eq:c-cut}
\end{equation}
where $M(\hat{\vec{n}})$ is a masking function (equal to $0$ or $1$ in
each pixel), so that the angular average denoted by the overbar is
over all point pairs (with $\hat{\vec{n}}_1 \cdot \hat{\vec{n}}_2 =
\cos \theta$) which lie outside the mask. This procedure is
mathematically identical to calculating the Legendre sum over the
pseudo-$C_{\ell}$ (PCL) estimates for the power spectrum (which we
denote $\hat{C}^{\mathrm{PCL}}_{\ell}$; see Appendix
\ref{sec:quadr-estim-some} for a precise definition):
\begin{equation}
\mathcal{C}(\theta)^{\mathrm{cut}} = \frac{1}{4\pi} \sum_\ell (2\ell+1) \hat{C}_\ell^{\mathrm{PCL}} P_\ell(\cos \theta) \textrm{.} \label{eq:c-cut-pcl}
\end{equation}
While this result has been reported before \cite{2004PhRvD..69h3524A},
an explicit proof does not appear to exist in the literature,
so we provide one in Appendix \ref{sec:estim-ctheta-s_12}.

It is clear that, if $\hat{C}_\ell$ are any unbiased estimates for
$C_\ell$, then forming their Legendre sum
(\ref{eq:c-full-leg}) yields an unbiased estimator for
$C(\theta)$. Thus if one wishes to find maximum likelihood estimates
for $C(\theta)$ on the full sky from cut sky information, by linearity
one simply substitutes the maximum likelihood $\hat{C}_{\ell}$
estimates in place of the PCL estimates. For our purposes, the
estimates provided by the quadratic maximum likelihood (QML) estimator 
technique \cite{1997PhRvD..55.5895T} are close enough to the exact
maximum likelihood to remove the need for any non-linear techniques
\cite{2004MNRAS.349..603E}.

There are two somewhat subtle points to be appreciated here.  First,
reservations have been expressed about the use of the QML estimator,
since it uses prior information on the power spectrum and therefore
appears to make strong assumptions about the form of the underlying
theory.  Copi {\it et al.} \cite{2008arXiv0808.3767C} express concern about
use of the QML estimator in circumstances where one is questioning the
validity of the concordance model. In fact, this unease turns out to
be unwarranted; one may explicitly show that QML estimates remain
superior to PCL estimates -- even in cases where the estimates for the
covariance matrix are not correct.

Let us outline why this should be so. A full derivation is given in
Appendix \ref{sec:relat-betw-qml}, in which the QML estimator is
written entirely within harmonic space. The estimation procedure can
be seen to down-weight high variance modes and up-weight low variance
modes before calculating the power spectrum of the weighted cut
sky. The resulting power spectrum is then correctly de-weighted and
de-convolved\footnote{This interpretation of the QML estimator's
  operation is a harmonic-space equivalent to the pixel-space
  `high-pass filter' interpretation given by Tegmark
  \cite{1997PhRvD..55.5895T}.}. The overall effect is to minimize the
cross-talk from the mask-induced mode-coupling.  Accordingly the QML
estimator will be close to optimal in reconstructing skies from any
theory where the power in each $\ell$ is close to that predicted in
concordance models. Since the observed power spectrum is very close to
the $\Lambda$CDM theoretical prediction, any serious candidate theory
must satisfy this criterion.  Indeed, only if the true theory has a
power spectrum closer to flat than to the concordance model will the
PCL estimator typically perform better than the QML. A corollary is
that the PCL and QML estimators become identical for a theory with
flat power spectrum\footnote{Closely related to this is the
  better-known result that PCL estimators for high $\ell$'s are nearly
  optimal for moderate sky cuts. Such cuts imply that the
  reconstruction is only sensitive to a finite window $\ell\pm \Delta
  \ell$, over which the high-$\ell$ $\Lambda$CDM power spectrum is
  nearly flat.} (equal $C_{\ell}$'s).

Efstathiou et al \cite{2009arXiv0911.5399E} make a different rebuttal
of the sensitivity-to-assumptions concern by showing that full-sky
maps (albeit band-limited) can be made from the cut sky data (see also
Ref. \cite{2006PhRvD..74b3005D}). The difficulty with full-sky
reconstructions is that they rapidly become unstable as the sky cut
increases (unless the data are strictly band-limited, which is not true
of the CMB). Furthermore they introduce a dependence on the underlying
theory so that their conceptual benefits over the QML estimator are
not clear (although, as shown in Ref. \cite{2009arXiv0911.5399E}, the
sensitivity to the assumed covariance may be rather weak). For this reason
we will not consider explicit sky reconstructions further in the
present work.

However the second subtlety is that one may not, in fact, want to
optimally reconstruct the full sky correlation function. If we are
interested in using a cut sky correlation function {\it not} to remove
localized contamination, but instead as a distinct quantity in its own
right, the efficiency of the QML estimator at reconstructing the full
sky becomes a hinderance. Starting instead from definition
(\ref{eq:c-cut}) -- the correlation of pixel pairs in a finite region
of the sky -- gives us a transparent interpretation. Therefore it
remains of interest to examine carefully the PCL-derived
$C(\theta)^{\mathrm{cut}}$, not because the QML estimator technique is
in doubt as a way of extracting reliable full-sky information, but
because the PCL technique explicitly extracts information which is
different.

% \begin{table*}
% \begin{tabular}{lllll}
% Notation & Interpretation & Equation  \\ \hline
% $\mathcal{C}(\theta)$ & Full-sky correlation function (from a single realization) \\
% $\mathcal{C}_\ell$ & Full-sky power spectrum (from a single realization) \\
% $\shalf$ & Measure of correlations on angular scales $>60^{\circ}$ (single realization) \\
% $C(\theta)$, $C_\ell$, $S_{1/2}$ & Ensemble averages of the above quantities \\
% $\hat{C}_\ell$ & Estimators for $C_\ell$ constructed from the cut sky & (\ref{eq:c-ell-est}) \\
% $\mathbf{C}$ & Covariance matrix for isotropic concordance theory & (\ref{eq:concordance-cov}) \\
% $\mathbf{C}'$ & Covariance matrix for underlying theory (when not concordance) & (\ref{eq:modified-cov}) \\
% $\mathbf{A}$ & Anomalous part of covariance matrix & $\mathbf{A} = \mathbf{C}'-\mathbf{C}$ \\
% $\mathbf{\Delta}^\ell$ & Identity matrix for $\ell$-harmonics; zero elsewhere & (\ref{eq:define-delta-l}) \\
% $\mathbf{R}^\ell$ & Reconstruction kernel & PCL case (\ref{eq:pcl-kernel}); QML case (\ref{eq:qml-kernel})\\
% $\mathbf{Z}^\ell$ & Reconstruction error kernel from full sky & $\mathbf{Z}^\ell = \tilde{\mathbf{R}}^\ell-\mathbf{\Delta}^\ell$
% \end{tabular}
% \caption{Quick reference for our notation. \app{Finish this table, or remove it}}
% \end{table*}

\section{Why is $\shalfcut$ is small?}\label{sec:why-mathc-small}

\begin{figure*}
\begin{center}
\includegraphics[width=0.99\textwidth]{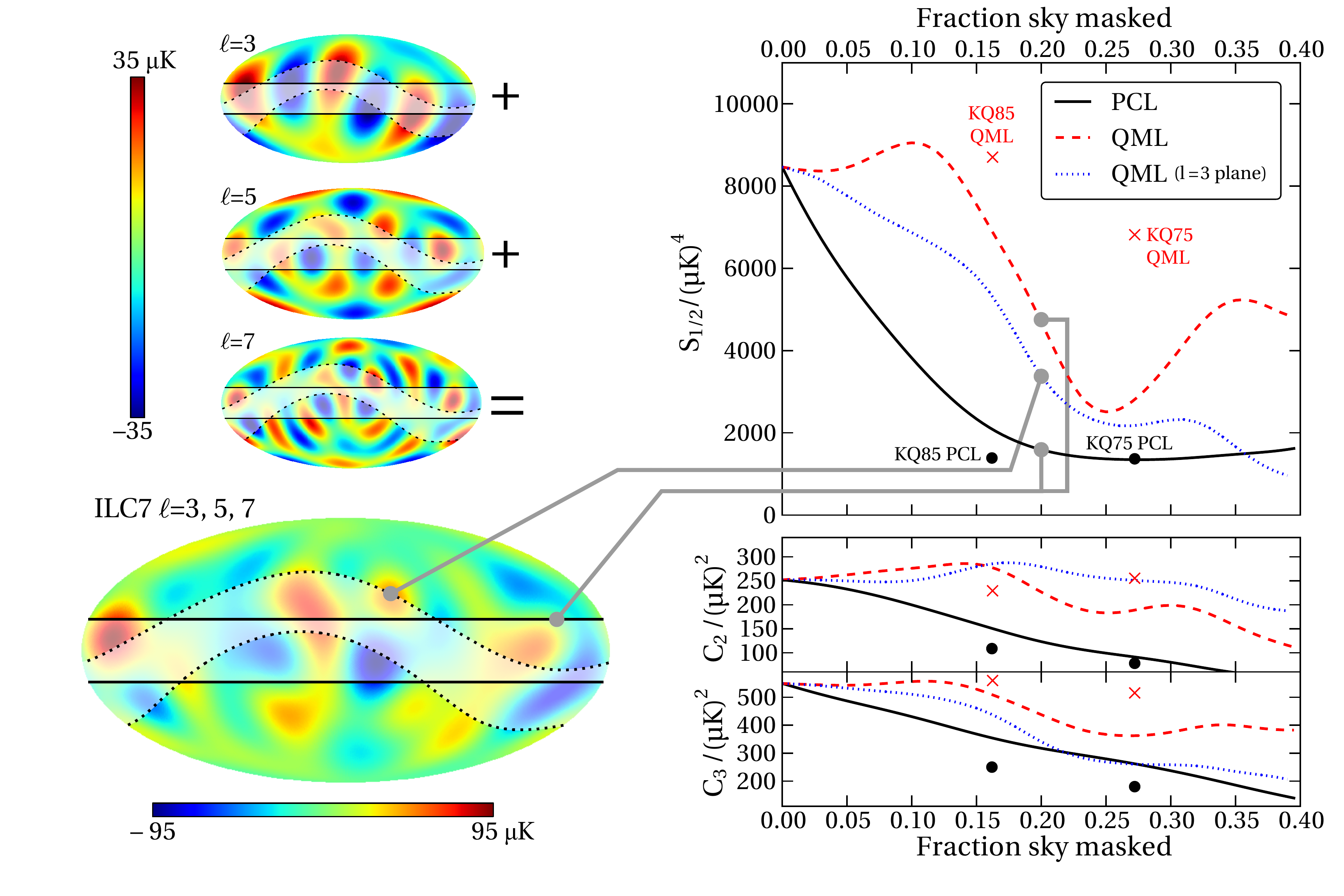}
\end{center}
\vspace{-0.5cm}
\caption{(Color online.)  An illustration of the low value of
  $\shalfcut$ and its origin. The upper left-hand panels show the
  pattern of ILC7 temperature fluctuations in $\ell=3, 5$ and $7$
  modes from top downwards; the larger panel underneath shows the sum,
  in which anticorrelations between the modes cause power near the
  poles to be small. The shaded regions bounded by solid and dashed
  lines represent a $20\%$ azimuthal mask in the Galactic and $\ell =3
  $ angular momentum planes respectively (defined in text). The large
  uppermost panel on the right hand side shows the values of
  $\shalfcut$ for a variable width Galactic azimuthal mask (solid
  line). The dashed line shows the same result using QML, rather than
  PCL, reconstruction techniques; the dotted line shows the QML result
  when the mask is applied in the $\ell=3$ angular momentum plane,
  demonstrating that full-sky power can be efficiently hidden even
  from the QML estimator. The two right-hand panels underneath the
  main plot show the corresponding estimated values $\hat{C}_2$ and
  $\hat{C}_3$. Plotted points show the results from using the WMAP
  team's KQ85y7 and KQ75y7 masks for PCL (circles) and QML (crosses)
  estimators respectively, exhibiting the QML estimator's relative
  sensitivity to mask shape.  }\label{fig:shalf-vary-cut}
\end{figure*}

In this section, we discuss the well-established result that
$\shalfcut$ is unexpectedly low (and much smaller than the full sky
value), and consider the origin of this observation from a harmonic
space perspective.  Known aspects of the full sky realization are
found to be behind the result, namely (a) the low amplitude, planarity
and rough Galactic alignment of the quadrupole; and (b) the planarity
and alignment of the octupole.

Combining equations (\ref{eq:shalf-quadratic}) and
(\ref{eq:c-cut-pcl}) shows that the value of $\shalfcut$ derived from
pixel-pair averages on the cut sky is:
\begin{equation}
\shalfcut = \sum_{\ell\ell'} \hat{C}^{\mathrm{PCL}}_{\ell} \hat{C}^{\mathrm{PCL}}_{\ell'} s_{\ell \ell'}\textrm{.}  \label{eq:shalf-cut-pcl}
\end{equation}
It is not clear why various groups have seen differences (albeit
minor) in $\shalfcut$ measured numerically on the sky and $\shalfcut$
defined by equation (\ref{eq:shalf-cut-pcl}), but it probably relates
to difficulties in designing stable numerical schemes which estimate
$\mathcal{C}(\theta)$ directly from pixel pairs.

\begin{figure}
\begin{center}
\includegraphics[width=0.5\textwidth]{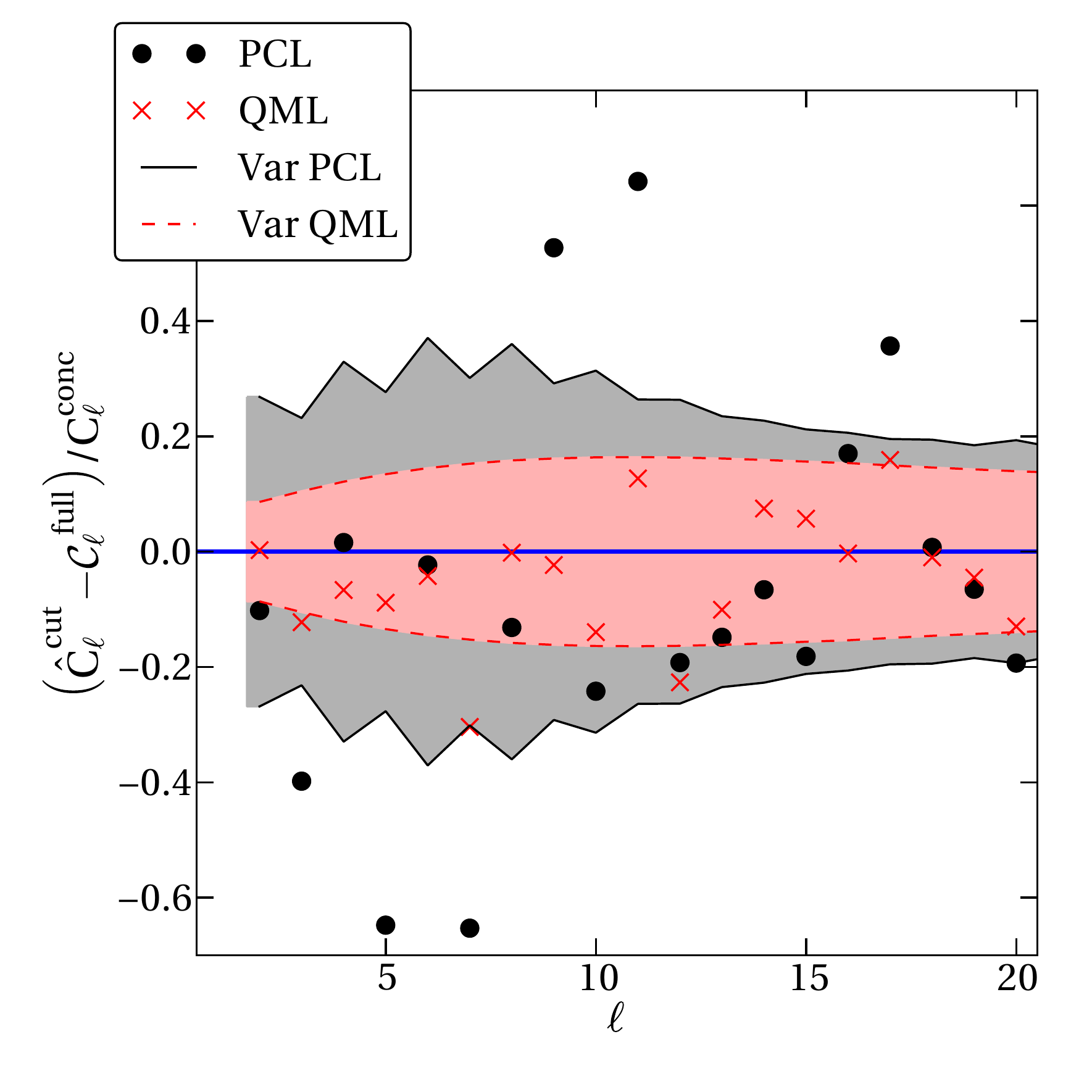}
\end{center}
\vspace{-0.5cm}
\caption{(Color online.) The difference between the cut-sky estimates
  $\hat{C}_\ell$ and the full-sky measured values $\mathcal{C}_\ell$
  on the ILC7 map, expressed for ease of viewing as a fraction of the
  fiducial theoretical best-fit $C_\ell$'s published by the WMAP team
  \cite{2010arXiv1001.4635L}. The results from a PCL and QML estimator
  for a $20\%$ azimuthal sky-cut are shown by dots and crosses
  respectively. The shaded bands show the expected deviation of the
  cut-sky from the full-sky values. The larger, outer band represents
  the PCL deviation while the smaller, inner band (with dotted edges)
  represents the QML deviation. The low value of $\shalfcut$ arises
  because the PCL power spectrum reconstruction for $l \leq 8$
  typically falls below, never significantly above, the full sky
  value.}\label{fig:c-ratios}

\end{figure}

To understand the origin of the observed $\shalfcut$ value, let us
first note that, in a typical realization of the isotropic
$\Lambda$CDM model, the primary driver of the ratio
$\shalfcut/\shalf^{\mathrm{full}}$ is the ratio
$\hat{C}_2^{\mathrm{PCL}}/\mathcal{C}_2^{\mathrm{full}}$. This
reflects the fact that, substituting the theory $C_{\ell}$'s in
equation (\ref{eq:shalf-quadratic}), the dominant contribution is from
the term quadratic in $C_2$.  However, due to the low amplitude of the
full sky quadrupole in our particular realization, the $\ell=2$ mode
becomes subdominant in determining $\shalf$. $C_3$ then
gives the dominant contribution on the full sky, so that
$\hat{C}_3^{\mathrm{PCL}}/\mathcal{C}_3^{\mathrm{full}}$
determines the magnitude of
$\shalfcut/\shalf^{\mathrm{full}}$.

Now the octupole of the observed realization happens to be somewhat
planar (although, as quantified below, not `anomalously' so). Its
preferred plane is, in turn, very roughly aligned with the Galactic
plane. When the octupole is masked using one of the standard WMAP
temperature analysis masks such as KQ75y7 (usable sky fraction
$f_\mathrm{sky}=70.6$\%) or KQ85y7 ($f_\mathrm{sky}=78.3$\%), a
significant amount of power in the octupole is hidden due to this
approximate alignment, leading the recovered
$\hat{C}_3^{\mathrm{PCL}}$ to be an under-estimate of the full sky
value. Thus, according to the considerations above, $\shalfcut$ drops
sharply in response.

Figure \ref{fig:shalf-vary-cut} illustrates this further by showing
(top plot, solid line) the value of $\shalfcut$ derived from the
seventh-year WMAP \cite{2010arXiv1001.4555G} Internal Linear
Combination map (ILC7) as the sky coverage of an equatorial, azimuthal
mask is increased from $0$ to $40\%$ ({\it i.e.} $f_{\mathrm{sky}}$
drops from $100\%$ to $60\%$). When the sky is unmasked, the PCL and
QML power spectrum reconstructions reduce to the full-sky estimate, so
that all results agree for the nil cut. For the moment we will focus
on the behaviour of the PCL reconstructions, returning to the QML
cases (dashed and dash-dotted lines) momentarily.

As described above, the rapid decline in $S_{1/2}^{\mathrm{cut}}$ as a
function of increasing mask width is largely due to the corresponding decline in
$\hat{C}_3^{\mathrm{PCL}}$ (illustrated in the lowermost right-hand
panel of Figure \ref{fig:shalf-vary-cut}), which in turn can be linked
to the progressive masking of the planar-concentrated power (see also
the $20\%$ sky-cuts illustrated in the CMB projections on the left of
the Figure). The known planarity of the quadrupole is also important,
in that the cut-sky estimates $\hat{C}_2^{\mathrm{PCL}}$ decline with
increasing mask area (central right-hand panel) and so do not regain
dominance over the octupole contribution.

The described properties of multipoles $\ell\le 3$ are not quite enough, on
their own, to account for the low $\shalfcut$.
Working with an azimuthal mask of $20\%$, if we use the cut sky values
$\hat{C}_{\ell}^{\mathrm{PCL}}$ for $\ell\le 3$ and the full sky
values $\mathcal{C}_{\ell}$ for $\ell>3$, we calculate a value of
$\shalf^{\mathrm{hybrid}}=3327\ \mu \mathrm{K}^4$. The true
cut-sky value for the same mask is $\shalfcut =
1529\ \mu \mathrm{K}^4$. We can account for this discrepancy by noting
that the PCL reconstructions of two other multipoles, $\ell=5$ and
$\ell=7$, are also rather low.

The overall situation is illustrated in Figure \ref{fig:c-ratios},
where we have plotted (as circles) the difference between full-sky and
cut-sky power spectra ($\hat{C}^{\mathrm{PCL}}_{\ell} -
\mathcal{C}_{\ell}$) at $f_{\mathrm{sky}}=80\%$; these are calculated
from the ILC7 map and scaled by the concordance theory $C_{\ell}$'s
\cite{2010arXiv1001.4635L}. The differences at $\ell=3, 5, 7$ are all
somewhat outside the $1\sigma$ variance (illustrated by the grey
jagged band)\footnote{The standard deviation illustrated in Figure
  \ref{fig:c-ratios} is defined as $\langle (\hat{C}_{\ell} -
  \mathcal{C}_{\ell})^2 \rangle^{1/2}$, {\it i.e.}  it is the
  `cut-induced' variance introduced in Appendix \ref{sec:expect-vari},
  equation (\ref{eq:civ}). We should note in passing that the variance
  is close to diagonal -- {\it i.e.} correlations between the
  estimates for different $\ell$ are small in both PCL and QML
  cases.}.

The reason for the shortfall in reconstructed power in $\ell=5$ and
$7$ is not immediately clear from inspecting their individual patterns
on the sky (small panels near top left of Figure
\ref{fig:shalf-vary-cut}). Only when all the odd multipoles at $\ell
\le 7$ are summed does the power become visually planar (see the
larger Mollweide projection at the bottom left of Figure
\ref{fig:shalf-vary-cut}). Thus cancellations between the $\ell=3,5,7$
modes in the polar regions effectively hide power from estimators once
the sky is masked. (We note that even-$\ell$ modes have no effect on
the odd-$\ell$ reconstruction and vice versa, since these are
decoupled when adopting an equatorially symmetric mask.)

\subsection{Behaviour of the QML estimator}

Having established the origin of the low $\shalfcut$, let us turn to
the effect of using a QML, rather than PCL, estimator in
reconstructing the full sky.  It has been commented elsewhere
\cite{2009arXiv0911.5399E} that, for KQ85y7 masks, the QML estimator
reconstructs most of the power in the full sky octupole;
$\shalf^{\mathrm{QML}}$ is close to
$\shalf^{\mathrm{full}}$ even for the larger KQ75y7 mask.
We have reproduced these results; QML outputs are plotted as crosses
in the panels of Figures \ref{fig:shalf-vary-cut} and
\ref{fig:c-ratios}. As discussed above (Section
\ref{sec:background-notation}), the QML estimator re-weights its input
maps to extract full-sky information as efficiently as possible; hence
the improvement is not surprising. Recall that, even when considering
anisotropic theories, the QML estimates for the full sky are expected
to be superior (this is further reinforced in Section
\ref{sec:theories} below).

However, the QML estimator's ability to make an efficient recovery of
our full sky results does depend on the shape of the mask\footnote{The
  PCL estimator is less sensitive to the exact shape of the cut than
  the QML estimator; this is to be expected given the simplicity of
  the former method (which, to a close approximation, measures the
  power on the cut sky and scales it by the appropriate sky
  fraction).}. For an azimuthal mask covering a sky fraction $\gtrsim
20 \%$, even the QML estimator starts to underestimate the power on
the full sky (illustrated by the dotted lines in the right panels of
Figure \ref{fig:shalf-vary-cut}; the lower panel shows that the
falling $\shalf^{\mathrm{QML}}$ tracks a drop in the power
of the reconstructed octupole $\hat{C}_3^{\mathrm{QML}}$). These
results show that, if the power is sufficiently localized within the
mask, it cannot be reconstructed by any technique.

This interpretation of the results is confirmed by applying an
azimuthal mask in the plane (as defined below) of the octupole (dotted
lines in all panels of Figure \ref{fig:shalf-vary-cut}). The octupole
plane is defined by rotating the map until the ``angular momentum
dispersion'' statistic \cite{2003PhRvD..68h3003D} for quantifying the
planarity of multipole $\ell$ is maximized:
\begin{equation}
L^2_{\ell} = \frac{\sum_{m= -\ell}^{\ell} m^2 |a_{\ell m}|^2}{\ell^2 \sum_{m=-\ell}^{\ell}  |a_{\ell m}|^2}\ . \label{eq:angmom-stat}
\end{equation}
(This maximization is achieved using a fast method described in
Appendix \ref{sec:rapid-calc-l2_m}.) Now the octupole is masked from
the map very efficiently, and $\hat{C}_3^{\mathrm{QML}}$ drops
sharply as a consequence. In response,
$\shalf^{\mathrm{QML}}$ becomes a severe under-estimate, at
large sky cuts becoming even worse than the Galactic azimuthal-masked
PCL estimator.

\subsection{Summary of the frequentist result}

The preceding material has shown that the small measured value of
$\shalfcut$ is attributable to a series of somewhat unlikely aspects
of the observed realization. We now recap and discuss briefly the
frequentist statistical significance of $\shalfcut$.

The primary contribution is the low ($254\,\mu\mathrm{K}^2$)
quadrupole amplitude (with a $P$-value of $4\%$ given the best-fit
power spectrum).  The planarity of the octupole can be assessed by
considering the rotation-maximized value of
Eq. (\ref{eq:angmom-stat}), which on the ILC7 map is 0.926. The
$P$-value computed from 10,000 isotropic realizations for observing
$L^2_3>0.926$ is $\sim$ 15\%, {\it i.e.} our realization is not
particularly unusual. The approximate alignment of this somewhat
planar octupole with the Galactic cuts typically used in CMB data
analyses can reasonably be regarded as purely coincidental ($P$-value
$21\%$). A consistent picture is found from assessing Figure
\ref{fig:c-ratios}, in which the $\ell=3$ PCL reconstruction deviates
from the mean by about $-1.5\sigma$. Similarly the shortfall of
PCL-reconstructed power in the $\ell=5$ and $7$ modes is a fluctuation
of around $-2\sigma$. None of these observations on their own look
particularly unusual; the statistical anomaly arises instead because
{\it all} of the low-$\ell$ PCL estimates are low.  Despite its
suboptimal nature the PCL estimator is unbiased, and the
reconstructions at different $\ell$'s are only weakly correlated, so
one would have expected as many over- as under-estimates.

The frequentist significance of the $\shalfcut$ result is connected,
then, to a series of coincident minor anomalies in our
realization. Only when combined in a specific way do these
observations raise frequentist alarm. Of course, this simply shows
that we have found a way to `factor' the low $P$-value of $\shalfcut$,
which does not by itself determine whether the anomaly might point to
theories beyond the concordance model.  Therefore in the next Section,
we consider the feasibility of finding theories which are
statistically preferred to the concordance theory in a Bayesian
comparison of $\shalfcut$.

\section{Anisotropic theories} \label{sec:theories}

\begin{figure*}
\begin{center}
\includegraphics[width=0.99\textwidth]{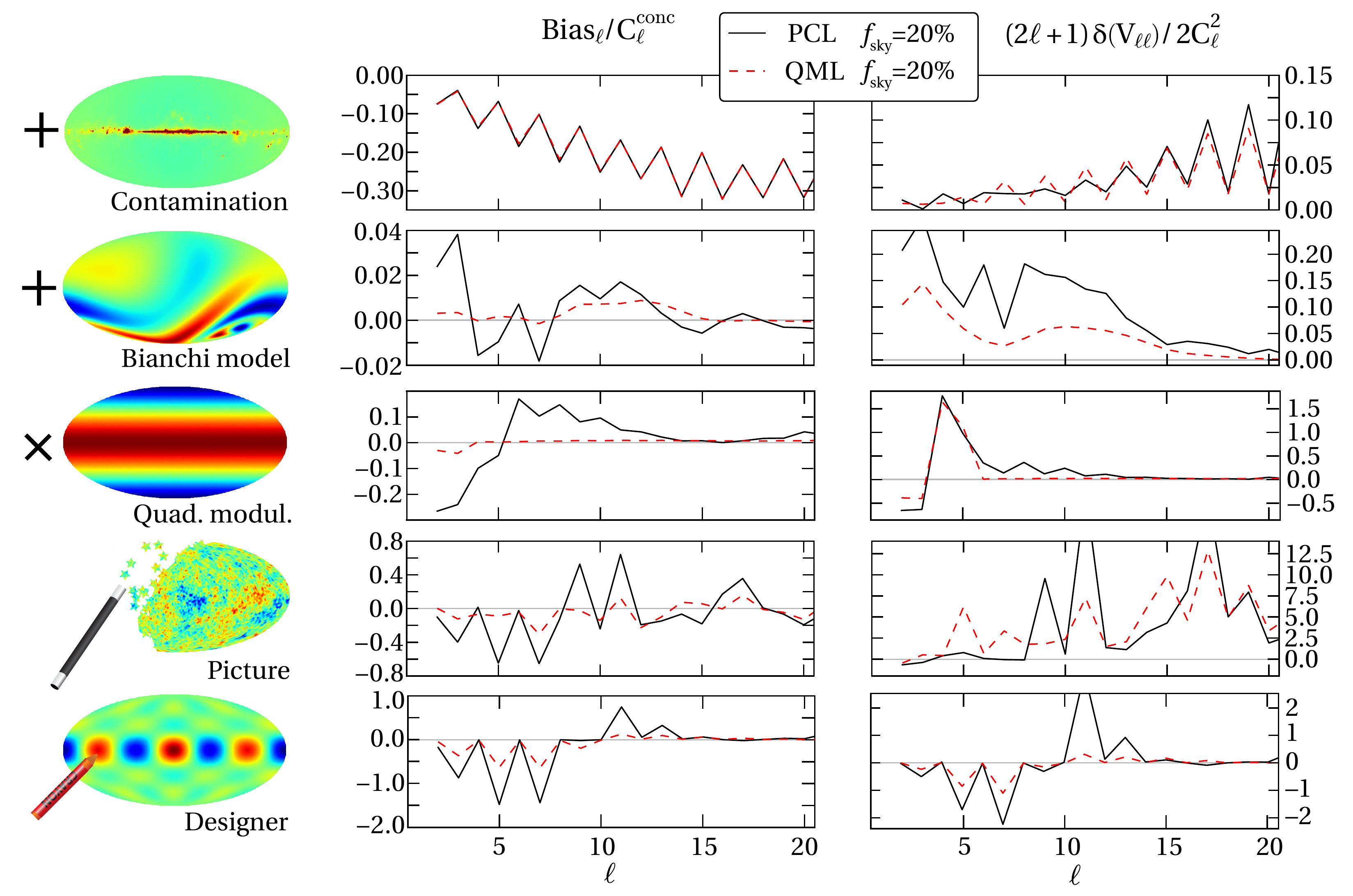}
\end{center}
\vspace{-0.5cm}
\caption{(Color online.)  For various anisotropic theories we plot the
  biases (left panels) and change in the diagonal part of the variance
  (right panels) compared to the isotropic case. In this context
  non-zero `bias' can be desirable, for instance in the case
  illustrated in the uppermost panels, where it merely reflects that
  the full sky power spectrum is contaminated by the presence of the
  galaxy. The solid lines show statistics for PCL reconstructions of a
  $f_{\mathrm{sky}}=20\%$ azimuthally masked sky; the dashed lines
  show the corresponding QML reconstructions. As expected, the QML
  reconstructions are generally more accurate (despite the anisotropy
  of the underlying theory); for more details see text. We scale the
  biases by the isotropic theory power spectrum and the change in the
  variances by the isotropic cosmic variance. The first three theories
  are simple existing models while the latter two are tests which are
  designed specially to reproduce the $\shalfcut$ anomaly. Note the
  different scales on each
  panel.}\label{fig:theories-biases-and-variances}
\vspace{-0.5cm}
\end{figure*}

%Having established a simple way to understand the origin of the low
%$S_{1/2}^{\mathrm{cut}}$ in harmonic space, we now turn to discuss
%whether such traits are to be expected in some form of anisotropic
%correction to the established theory of the primary CMB. 

We have shown in the previous Section that the low observed
$S^{\mathrm{cut}}_{1/2}$ can be attributed to the fortuitous alignment
of power in the $\ell=3, 5$ and $7$ modes of the CMB (along with the
planarity, and small full-sky amplitude, of the quadrupole).

Broadly, one can imagine three distinct ways in which the small
observed cut-sky power in the $\ell=3, 5$ and $7$ modes could look
less anomalous in an alternative theory:

\begin{enumerate}
\item The $\hat{C}^{\mathrm{PCL}}_{\ell}$ estimates could turn out to
  be biased in the ensemble mean of the true theory;
\item The $\hat{C}^{\mathrm{PCL}}_{\ell}$ estimates could have a
  larger variance in the ensemble of the true theory, making the
  departures from the mean less significant;
\item The true theory could correlate $\hat{C}^{\mathrm{PCL}}_{\ell}$
    estimates so that the likelihood of small cut-sky power in
    $\ell=5, 7$ is greater once the small cut-sky power in $\ell=3$ is
    known.
\end{enumerate}
Our main focus in what follows will be on (1); possibilities (2) and
(3) will be mentioned where relevant. 

Given a true covariance matrix $\mathbf{C}$ related to the concordance
isotropic theory by
$\mathbf{C}=\mathbf{C}^{\mathrm{conc}}+\mathbf{A}$, one may explicitly
calculate the bias,
\begin{equation}
\mathrm{Bias}_{\ell} = \langle \hat{C}_{\ell} - \mathcal{C}_{\ell} \rangle\label{eq:aniso-bias}
\end{equation}
and the variance
\begin{equation}
V_{\ell\ell'} = \left\langle \left( \hat{C}_{\ell} - \langle \hat{C}_{\ell} \rangle \right)\left( \hat{C}_{\ell'} - \langle \hat{C}_{\ell'} \rangle \right) \right \rangle\textrm{,}\label{eq:aniso-variance}
\end{equation}
where the false covariance matrix $\mathbf{C}^{\mathrm{conc}}$ is used
in constructing estimators, but the true covariance matrix
$\mathbf{C}$ is employed in taking the final ensemble average. For
algebraic expressions the reader is referred to Appendix
\ref{sec:expect-vari}.  

% The bias has been defined as the expected difference between cut-sky
% and full-sky measured power spectra. This allows it to be calculated
% in a uniform way for each underlying theory. However, by this
% definition, we may not always want the bias to be zero, since some
% specific theories allow more physical distinctions between `desirable'
% and `undesirable' contributions to the reconstructed power spectrum --
% a good example is the Galactic contamination example described below.

To gain a feel for how alternative theories can influence the
recovered power spectrum on the cut sky, let us consider the following
specific cases.

\renewcommand{\labelenumi}{(\roman{enumi})}
\begin{enumerate}
\item {\it Galactic contamination}, {\it i.e.} residual errors in the
  Galactic signal subtraction. This is modelled by creating a template
  map of possible errors, taking $1\%$ of the difference of the WMAP7
  ILC map with the V-band map (after smoothing to a common resolution
  of $1^{\circ}$). The map gives us a rough handle on the form of the
  residual contamination to be expected (albeit with an unknown
  amplitude).  In the ensemble, the template map is simply added to
  the observed CMB sky, yielding equivalent results to a theory with
  anisotropic Gaussian correction
  $\mathbf{A}=\vec{g}\vec{g}^{\dagger}$, where $\vec{g}$ represents
  the spherical harmonic coefficients of the contamination map.  It
  has already been shown in Ref. \cite{2008PhRvD..78l3509B} that this
  kind of contamination cannot improve the likelihood of $\shalfcut$,
  but the model remains helpful for our discussion below.
%\item ??{\it Hot pixel contamination}, another form of residual error in
%  the map-making process put forward by \app{ref 0903.3133} as
%  relevant to the $S_{1/2}$ anomaly. We make a toy model of this type
%  of contamination in the V-band by simulating the WMAP scan strategy
%  with an uncorrected $1\%$ common mode response. After creating a map
%  from our simulated time ordered stream, we subtract the input map to
%  create a contamination template which is treated in the same way
%  as the galactic template above. \app{Too complex, remove?}

\item {\it Bianchi VII$_h$ template}. Using the algorithms of Refs.
  \cite{2007MNRAS.380.1387P,PC08} we calculate a temperature
  anisotropy template for the Bianchi VII$_h$ vector mode case with an
  amplitude of $35\,\mu$K according to the best-fit parameters of
  Ref. \cite{2006ApJ...643..616J}. In the ensemble this is added to
  the concordance CMB as with the templates considered
  above. Physically, such a setup can be motivated by the existence of
  anisotropic Bianchi modes which are well behaved at the initial
  singularity, although such models are fine-tuned.

\item {\it Quadrupolar Modulation}. A strong quadrupolar
  modulation\footnote{ Attention has also been given in the past to
    dipolar modulations.  In the case of an equatorial azimuthal mask
    this can have only second order effects on power spectrum
    reconstruction, since it couples $\ell$ to $\ell\pm 1$ while the
    mask couples $\ell$ to $\ell\pm 2n$. A different quadrupolar
    anisotropy, that of the inferred primordial power spectrum, has
    also been reported
    \cite{2009ApJ...690.1807G,2009PhRvD..80f3004H,2009arXiv0911.0150G}. However,
    Hanson et al. \cite{2010arXiv1003.0198H} have identified WMAP beam
    asymmetries as the origin of this unconnected effect.} of the
  temperature field is known to reproduce the co-planarity of
  quadrupole and octupole \cite{2005PhRvD..72j3002G}. The modulation
  is required to have a very large amplitude, yet be confined to low
  multipoles. Dvorkin {\it et al.}  \cite{2008PhRvD..77f3008D} discuss
  how any early universe model of such a modulation must be carefully
  tuned in harmonic space to avoid the leakage of modulated power to
  high multipoles through projection effects.  We approximate these
  considerations by modulating only the quadrupole and octupole of
  isotropic realizations. Since a quadrupolar modulation of a
  multipole $\ell$ couples power to $\ell\pm2$, on the full sky our
  modulation only has an effect on multipoles $\ell\le 5$.
  
\item {\it Picture} and {\it Designer} theories. These are specific
  theories designed to investigate the best possible statistical gains
  to be made from anisotropic theories over the concordance case. We
  will describe them in detail in Section
  \ref{sec:designer-theory-main}, below.

\end{enumerate}

The left panel of Figure \ref{fig:theories-biases-and-variances}
exhibits the biases induced by each of these theories [defined by
equation (\ref{eq:aniso-bias})]; the right hand panel shows the
diagonal part of the variance [defined by equation
(\ref{eq:aniso-variance})]. In both panels, the results from the PCL
estimators are plotted as solid lines, while the QML results are shown
by dashed lines. The sky cut imposed for these calculations is a
$20\%$ Galactic azimuthal mask.

In the first case, that of Galactic contamination, the `bias' reflects
the added power from the Galaxy, visible in the full sky
$\mathcal{C}_{\ell}$'s but naturally invisible to reconstructions made
from the cut sky (in which the Galaxy is masked away). Both the PCL
and QML estimators therefore become equally `biased', but this is, in
fact, a desirable feature: they are rejecting the contamination. Note
that the zig-zag pattern in the bias arises from the rough equatorial
symmetry of the Galaxy, which results in a much stronger coupling to
even, rather than odd, $\ell$'s. On the other hand the zig-zag in the
biases actually observed (Figure \ref{fig:c-ratios}) is larger at odd,
rather than even, $\ell$'s. Finally, the apparent biases and extra
variance of spatially localized contamination tends to grow towards
high $\ell$ as a fraction of the full sky power, whereas the observed
discrepancies are confined to low $\ell$.

The second case (Bianchi contamination) is similar in that it adds a
template to the concordance covariance; but because the power is not
localized within the mask, it is now visible even on the cut-sky. As
expected from our earlier considerations, the QML estimator in this
regime reconstructs the full sky $\mathcal{C}_{\ell}$ power with a
smaller bias and variance than the PCL case. The increased variance
(of order $20\%$ of the cosmic variance) is more significant than the
bias (of order $2\%$ of the power spectrum). This can be understood by
noting that, since the Bianchi signal has a small rms power of
$\sim 12\, \mu$K$^2$, the individual elements of the template
covariance contribution $\mathbf{A}$ are much smaller than the
elements of the concordance covariance matrix
$\mathbf{C}^{\mathrm{conc}}$. Expanding an expression for the variance
(\ref{eq:estimator-variance}) highlights the existence of cross-terms
in $\mathbf{C}^{\mathrm{conc}}$ and $\mathbf{A}$; it is these leading
order contributions which give the larger variance.

Let us now turn away from additive modifications to the concordance
theory, and instead discuss the quadrupolar modulation.  When analysed
on the cut sky, the power at low multipoles is hidden from the PCL
estimator (but less so from the QML estimator), leading to a negative
bias.  The modulation also couples $\ell \pm 2$, creating power on the
full sky in $\ell=4, 5$; this accounts for the spike at these
multipoles in the variance of the estimators. The extra power is
further spread to higher $\ell$ by mode-coupling resulting from the
masking operation. The result is that the PCL estimator over-estimates
power in multipoles $5<\ell<15$; note that, because the power spectrum
is rapidly decreasing, a small leakage of power to high $\ell$ from
the low multipoles can result in a substantial bias. Once again the
QML estimator fares better, more efficiently confining the
contamination to low $\ell$.

The quadrupolar modulation behaves qualitatively as expected, hiding
power at low $\ell$; this is the right sort of effect to reproduce the
low $\shalfcut$ and therefore produces a small increase in the
likelihood of the observed value. However, Figure
\ref{fig:theories-biases-and-variances} shows that the biases from
this theory are rather small. Therefore, rather than focus on this
model, we can go one stage further and consider tuning Gaussian models
to fit the value of $\shalfcut$ as closely as possible.  These
theories are less transparent in their physical meaning, but are
guaranteed to give a better fit to the observed properties of the sky.

\subsection{Picture theory}

Consider the theory which exactly matches the observed CMB; it has
covariance matrix $\mathbf{C}=\vec{a}\vec{a}^\dagger$ where $\vec{a}$
represents the observed ILC $a_{\ell m}$'s. Because $\mathbf{C}$ has
zero variance in any direction orthogonal to the observed data, it has
an infinite likelihood (or, more correctly, a likelihood bounded from
above only by noise in the experiment).

The ensemble for this theory is a series of pictures of our own CMB
sky (represented here by the ILC map), scaled by a Gaussian random
amplitude of unit variance. Consequently the biases exactly match the
values of $\hat{C}_{\ell} - \mathcal{C}_{\ell}$ for our observed sky
(see Figure \ref{fig:theories-biases-and-variances}). However, because
there is only one mode (the amplitude scaling of the entire sky), the
variances become extremely large. To build intuition, consider the
cosmic variance of the concordance model, in which the variance on
$\mathcal{C}_{\ell}$ decreases as $2\ell +1$. This arises solely because
of the additional modes available at increasing $\ell$; in the picture
theory all modes are perfectly correlated, so the cosmic variance does
not decline in this way.

In spite of the divergently large likelihood for the picture theory,
the variance means that our observed value of $\shalfcut$ has a finite
likelihood which can be calculated by Monte Carlo simulation of the
ensemble. In Figure \ref{fig:shalf-likelihoods} we plot the log
likelihood for the concordance $\Lambda$CDM model (solid line), the
picture theory (dash-dotted line) and the designer theory (dashed
line), the last of which we will return to momentarily. 

The improvement in the log likelihood of $\shalfcut$ for the picture
model (over the isotropic concordance case) is $\Delta \ln \mathcal{L}
= 3.7$.  This disappointingly modest improvement can be seen to result
from the large cosmic variance in a theory with only one degree of
freedom: while it peaks near the observed value, the $\shalfcut$
likelihood function for the picture theory is extremely broad. The
broadness in turn impacts upon the peak value because the likelihood
must be normalized to one.

\subsection{Designer theory}\label{sec:designer-theory-main}

We have examined the picture theory in which the likelihood of our own
observed CMB is, by design, divergently large. The $\shalfcut$
likelihood was shown to barely favour the theory because the variance
on this value is so large. In this section we search for a theory with
similar properties to the observed sky, but allow power to be spread
through many more modes, so that the variance in high-order statistics
such as $\shalfcut$ is better controlled. 

\begin{figure}
\begin{center}
\includegraphics[width=0.49\textwidth]{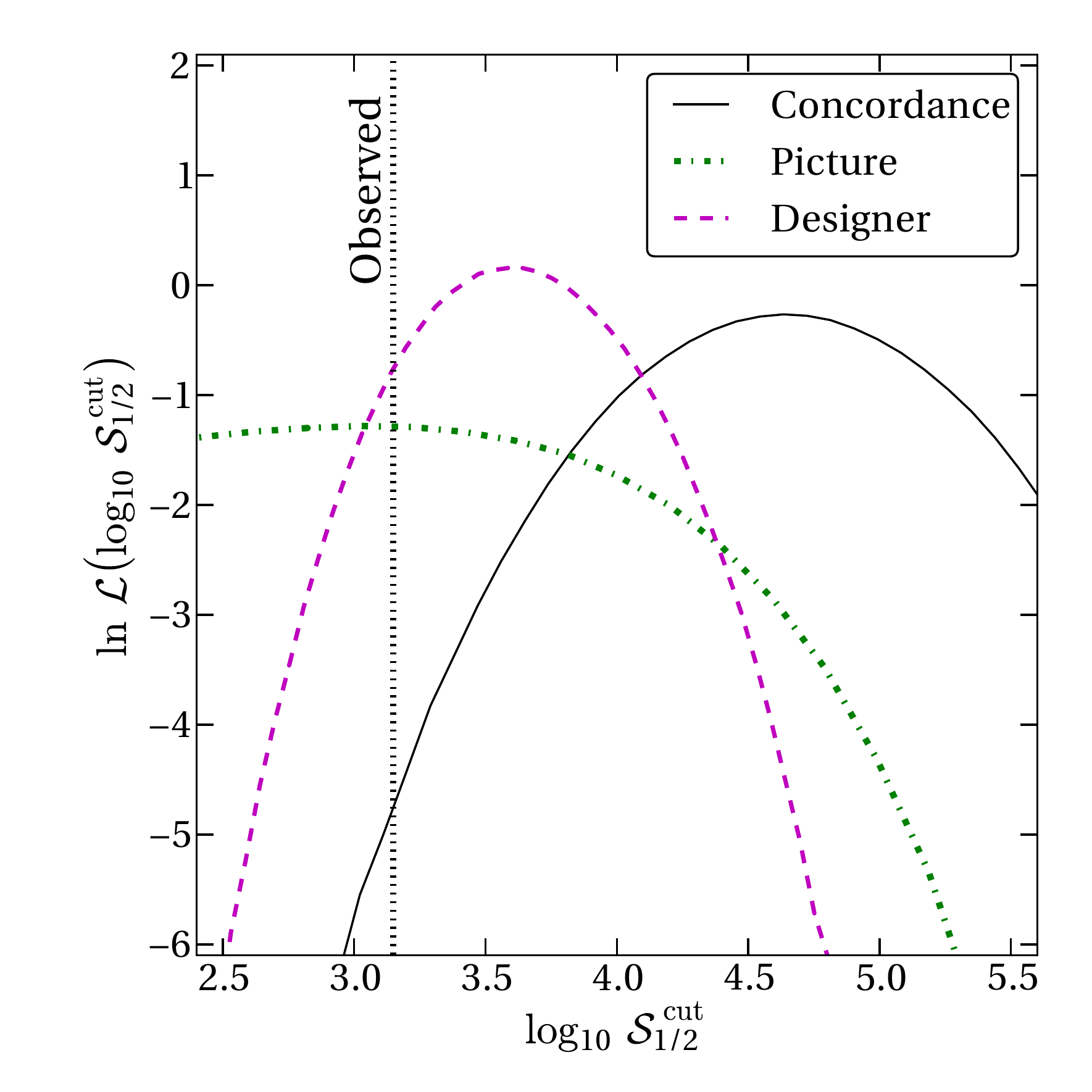}
\end{center}
\vspace{-0.5cm}
\caption{(Color online.) The likelihoods for $\shalfcut$ compared
  between three theories: the isotropic concordance theory (solid
  line), the `designer' theory (dashed line) and `picture'
  (dash-dotted line) theory. The latter two are specifically designed
  to reproduce low $\shalfcut$. The improvement in the log likelihood
  over the concordance cases are respectively 4.2 and 3.7, which are
  very small improvements given the fine-tuning involved.  Since these
  theories should produce the greatest possible gains in likelihood,
  the value of the observed $\shalfcut$ statistic is not a strong
  objection to the concordance theory.}\label{fig:shalf-likelihoods}
\end{figure}

A full explanation of the mathematical construction is given in
Appendix \ref{sec:designer-theories}. Our search is over all positive
definite covariance matrices $\mathbf{C}$, corresponding to all
Gaussian theories with zero mean ($\langle \Delta T \rangle = 0$),
subject to two sets of constraints. Firstly, the covariance matrix is
required to have full-sky theory $C_{\ell}$'s [defined by
Eq. (\ref{eq:theory-cl-definition})] equal to the observed ILC values
($C_{\ell}=\mathcal{C}^{\mathrm{ILC}}_{\ell}$). Secondly, the PCL
estimator applied to the theory on the $20\%$ azimuthally masked sky
is required to find zero power in $\ell=2, 3, 5$ and $7$ ({\it i.e.}
$\mathrm{Bias}_{\ell} = -\mathcal{C}_{\ell}$ at these $\ell$'s).  In
order to satisfy these results simultaneously, our technique naturally
introduces anisotropic correlations between different multipoles.
Substantial freedom remains, which we use to minimize the cosmic
variance of the final theory (see Appendix
\ref{sec:designer-theories}).  The freedom is truncated at
$\ell_{\mathrm{max}} = 10$; we adopted the concordance
covariance for all $\ell>\ell_{\mathrm{max}}$, but verified
that our conclusions are insensitive to this choice.

The CMB projection labelled `Designer' in Figure
\ref{fig:theories-biases-and-variances} illustrates an actual
realization from this model (although only the $\ell=3, 5$ and $7$
modes are plotted). One can see that the theory is very efficient at
localizing power in modes confined within our specified $20\%$ mask
(while keeping the full sky power spectrum to the specified values).
The bias panel for this theory shows that, accordingly, no power is
detected by the cut-sky PCL estimator in $\ell=3, 5$ and $7$. (The plot
shows that some of the biases are actually smaller than $-C_{\ell}$,
which is as expected since
$\mathcal{C}^{\mathrm{ILC}}_{\ell}>C_{\ell}$ at the corresponding
multipoles.)  At low $\ell$, the variance on the $\hat{C}_{\ell}$'s is
smaller than for the concordance model ($\delta V_{\ell\ell} <0$),
because the reconstructed power in these modes remains close to zero
in all realizations.  For $\ell\gtrsim 10$, there is a spike of larger
variance arising from the mask-induced contamination similar to that
described for the modulation model. At large $\ell$, the variance
tends to the standard concordance variance for the estimators ($\delta
V_{\ell\ell}=0$).  Once again, the QML estimator performs better in
minimizing both bias and variance compared to the PCL case at nearly
all $\ell$.

We can now return to Figure \ref{fig:shalf-likelihoods} which displays,
as a dashed curve, the likelihoods for our designer theory. We
described above how the power localization at low $\ell$ favours a low
$\shalfcut$; accordingly the peak likelihood (at
$\log_{10}\,\shalfcut/\mu\mathrm{K}^4 = 3.6$) is considerably smaller
than the equivalent value for the concordance theory
($\log_{10}\,\shalfcut/\mu\mathrm{K}^4 = 4.6$). However, despite being
minimized by spreading power through more degrees of freedom, the
variance of the designer $\shalfcut$ remains large and consequently
the improvement in likelihood is modest ($\Delta \ln \mathcal{L} =
4.2$) despite the dramatic increase in the number of degrees of
freedom needed to construct this theory.

It is clear from Figure \ref{fig:shalf-likelihoods} that to obtain
significant gains in likelihood for $\shalfcut$, one needs to achieve
far smaller cosmic variance on $\shalfcut$. But the designer theory
plausibly gives near the smallest possible variance on this
quantity. In particular, Appendix \ref{sec:designer-theories} derives
a minimum bound on the variance of $\shalfcut$. The lower bound can be
understood as arising from a suitable isotropic limit (which is
unattainable in practice, but provides a provable lower limit for
attainable theories); isotropic theories minimize the cosmic variance
for a given power spectrum, because they maximize the number of
independent modes and spread power through these modes as evenly as
possible. The lower bound calculated from the appropriate isotropic
test-case is $\sigma^2_{\mathrm{min}}(\shalfcut) = 1.7 \times 10^7 \mu
\mathrm{K}^8$, compared with the variance on the designer theory of
$\sigma^2(\shalfcut)=2.9 \times 10^7 \mu \mathrm{K}^8$. Thus the
designer theory detailed in this section almost saturates the variance
limit; we may be confident that no Gaussian theory can have a
significantly more peaked likelihood.

\subsection{Summary and discussion}\label{sec:summary-discussion}
\vspace{-0.1cm}

The two theories we have discussed in the preceding two sections (the
first giving an infinite likelihood for the observed sky; the second
tuned as far as we can to give a large likelihood for the single value
$\shalfcut$) strongly suggest that no anisotropic Gaussian theory can
improve the likelihood of the observed $\shalfcut$ by more than\linebreak
$\Delta \ln \mathcal{L} \simeq 5$. Because of the careful fine-tuning
of these models, they form a plausible upper bound for the statistical
gain available. 

How can we interpret this very modest likelihood gain?  From a
Bayesian perspective, correlations between the primary temperature
(from the high-$z$ last scattering surface) and the integrated
Sachs-Wolfe signal (from local structure) are implied in any model
aligning low-$\ell$ power
\cite{2008PhRvD..77f3008D,2009arXiv0909.2495F,2009arXiv0911.5399E}. Therefore any
realistic physical prior probability is very small relative to the
isotropic $\Lambda$CDM cosmogony; the posterior probability ratios
will still vastly favour the latter theory.  

An alternative argument is as follows. Up to $\ell_{\mathrm{max}}=10$
(and excluding monopole and dipole), the anisotropic theory has
approximately $6\,900$ degrees of freedom compared to the isotropic
case with $8$ degrees of freedom. Thus the improvement in log
likelihood per degree of freedom is of order $10^{-3}$. While this is
not a strictly Bayesian interpretation, it does suggest that our
statistical gain has been achieved only with enormous fine-tuning.

Both of these lines of reasoning suggest strongly that the observed
value of $\shalfcut$ can never constitute strong evidence in favour of
any Gaussian theory posited as an alternative to $\Lambda$CDM.
One escape route from this result is to consider, rather than the
absolute value of $\shalfcut$, the ratio $\shalfcut/\shalf$. In the
case of the picture theory, for instance, one then has an infinite
likelihood for the observed value of the ratio (because the one
available random degree of freedom -- the amplitude -- cancels between
numerator and denominator). This immediately demonstrates that there
is no upper bound to the likelihood gain for such a statistic.  We
would suggest, however, that the existence of an upper bound for the
original statistic, $\shalfcut$, is an attractive property --
precisely because it allows for an understanding of the Bayesian
theoretical improvements available without detailed physical
modelling. The most convincing way to show that the observed sky is
anomalous would therefore be to find a statistic encapsulating the
planarity and correlation of power which has large, but not trivially
infinite, likelihood gains available. Starting from the results of
Section \ref{sec:why-mathc-small}, such a statistic might be developed
from physical considerations on the full sky.

\vspace{-0.2cm}
\section{Conclusions}\label{sec:conclusions}
\vspace{-0.2cm}
There is a classic difficulty in understanding large and complex
datasets such as those produced by WMAP and, in the future, {\it
  Planck}: they contain so much information that statistical anomalies
can be found without any difficulty. We have taken as an example the
purported anomalous aspects of the angular correlation function.  Some
previous work claims that, after considering these anomalies, the
entire cosmological paradigm is to be doubted
\cite{2008arXiv0808.3767C}; other authors claim that apparent
anomalies can be dismissed as the product of {\it a posteriori}
analysis \cite{2009arXiv0911.5399E}. Yet {\it a posteriori} reasoning
must be allowed in science, since otherwise we would rarely, if ever,
recognize failings of our existing knowledge.

The contrary statistical claims relating to $S_{1/2}^{\mathrm{cut}}$
are reconciled by appreciating that, without an alternative theory to
test against, there is no unambiguous significance to any anomaly. We
have therefore presented an alternative approach to this puzzle: we
examined the origin of the low $S_{1/2}^{\mathrm{cut}}$ in harmonic
space, and then attempted to find theories that reproduce the required
patterns.

In the process we noted that the cut-sky correlation function contains
identical information to the PCL power spectrum estimates. We
therefore used the PCL estimates for the majority of our results, but
also demonstrated that the standard QML techniques provide more
reliable reconstructions of the full sky, even when anisotropy is
suspected.  We informed our intuition about the behaviour of the
estimators by considering simple anisotropic modifications to the
concordance models (contamination, Bianchi and quadrupolar modulation
theories). This showed explicitly that the QML estimator biases
introduced by anisotropic theories were smaller than or comparable to
the PCL case.

Then, by attempting to construct anisotropic Gaussian theories which
improve the likelihood of the low $\shalfcut$, we demonstrated that no
significant gains in likelihood for this single statistic are
available.  Since there is no suggestion in the observed sky that the
underlying ensemble is significantly non-Gaussian
\cite{2010arXiv1001.4538K}, it is implausible that post-Gaussian
corrections would substantially change our results. We therefore
conclude that the $\shalfcut$ anomaly is not likely to point to new
physics. 

If it does have any meaning, the $\shalfcut$ anomaly (and the
underlying shortfall of power seen by PCL estimators) does not
indicate a vanishing large-scale correlation function, but rather is
related to alignments of low-$\ell$ power on the full sky (Section
\ref{sec:why-mathc-small}).  It is likely that full-sky statistics can
be constructed which capture these unexpected correlations better than
$\shalfcut$ -- and these could evade our likelihood limits.  However,
we argued that more trivial modifications (such as taking the ratio
$\shalfcut/\shalf$) which sidestep our constraint by attaining an
infinite likelihood under the `picture' theory
($\mathbf{C}=\vec{a}\vec{a}^{\dagger}$) are not helpful; see Section
\ref{sec:summary-discussion}. In other words it is highly desirable to
choose statistics, such as $\shalfcut$, that do allow for a finite
limit to be placed on the Bayesian statistical gain available under a
wide class of alternative straw-man models.  Considering the magnitude
of that limit is then, in our view, a plausible way to probe the
significance of {\it a posteriori} anomalies.

\vspace{-0.4cm}
\subsection*{Acknowledgments}
\vspace{-0.3cm}
We thank Anthony Challinor, Steven Gratton, Daniel
  Mortlock, Antony Lewis, Wayne Hu, George Efstathiou, Glenn Starkman,
  Dragan Huterer and Dominik J. Schwarz for productive discussions.
  AP is supported by Emmanuel College, Cambridge.  HVP is supported by
  Marie Curie grant MIRG-CT-2007-203314 from the European Commission,
  and by STFC and the Leverhulme Trust.  HVP thanks the Aspen Center
  for Physics for hospitality.  We acknowledge use of the Healpix
  package \cite{2005ApJ...622..759G} and the Legacy Archive for
  Microwave Background Data Analysis (LAMBDA).  Support for LAMBDA is
  provided by the NASA Office of Space Science.

\appendix

\section{Quadratic Estimators: some useful results}\label{sec:quadr-estim-some}

In this Appendix, we summarize some technical details omitted from
the main paper alongside useful results pertaining to the two
most common quadratic power spectrum estimators: the pseudo-$C_{\ell}$
and quadratic maximum likelihood estimators. These will be introduced
and compared in a single quadratic estimator framework to gain
insights into their similarities and differences. Since we work in
harmonic space, we first explain the sky-masking operation.

\subsection{Cutting the sky}\label{sec:cut-sky}

It is standard practice in CMB analysis to remove regions of the sky
in which contamination from the Galaxy (or other undesirable sources)
is suspected. This is accomplished by masking the temperature field
and then constructing measurements based solely on the masked
data. In harmonic space, the masked temperature expansion coefficients
$\tilde{a}_{\ell m}$ are related to the unmasked $a_{\ell m}$ via
\begin{eqnarray}
\tilde{a}_{\ell m} & = & K_{\ell m\ell' m'} a_{\ell 'm'}\textrm{, where} \label{eq:masked-alm}\\
K_{\ell m\ell' m'} & = & \int \dd \Omega \, Y^*_{\ell m}(\Omega) Y_{\ell 'm'}(\Omega) M(\Omega) \textrm{.} \label{eq:mask-matrix-def}
\end{eqnarray}
Here $M(\Omega)$ is $0$ within the masked region and $1$
outside. [See Appendix \ref{sec:harmonic-space-masking} for a brief
discussion of a hidden numerical pitfall in equation
(\ref{eq:mask-matrix-def}).]  We can write expression
(\ref{eq:masked-alm}) compactly as the linear transformation
\begin{equation}
\tilde{\vec{a}}  =  \mathbf{K} \vec{a} \textrm{,}
\end{equation}
where $\vec{a}$ is a
vector composed of the $a_{\ell m}$'s. 

It will be helpful to note that $\mathbf{K}$ is both
idempotent ($\mathbf{K}^2=\mathbf{K}$) and Hermitian
($\mathbf{K}^{\dagger}=\mathbf{K}$). These identities may both be
derived straight-forwardly from equation~(\ref{eq:mask-matrix-def});
together they allow much flexibility in manipulating certain
equations.
In these appendices, the addition of a tilde will represent masked
quantities and operators; thus $\tilde{\vec{b}}=\mathbf{K}\vec{b}$ for
any data vector $\vec{b}$, while for any matrix $\mathbf{M}$ we write
\begin{equation}
\tilde{\mathbf{M}} = \mathbf{K} \mathbf{M} \mathbf{K}\textrm{,} \label{eq:cut-sky-conj}
\end{equation}
implicitly taking advantage of the Hermitian property. For most of our
numerical results we have assumed the mask is azimuthally symmetric,
$M(\theta,\phi)=M(\phi)$. This is a reasonable approximation to true
Galactic masks, and results in enormous computational simplification
because $\mathbf{K}$ becomes sparse,
\begin{equation}
K_{\ell m\ell' m'} = K^m_{\ell\ell'} \delta_{mm'} \textrm{ (no sum).}\label{eq:azimuthal-simplification}
\end{equation}
However all algebraic results are obtained with no such assumptions
and are applicable to any type of mask.

Since the estimators considered here are quadratic in the cut-sky
$\tilde{a}_{\ell m}$'s, we may write for a generic estimate
$\hat{C}_{\ell}$:
\begin{equation}
\hat{C}_{\ell} = \tilde{\vec{a}}^{\dagger} \mathbf{R}^{\ell} \tilde{\vec{a}} \label{eq:c-ell-est}
\end{equation}
for some set of matrices $\mathbf{R}^\ell$. Before explicitly defining
these matrices, we describe a helpful notational trick and discuss a
couple of generic features of quadratic estimators.

\subsection{A helpful notational trick}

Recall that in Sec. \ref{sec:background-notation} the power spectrum
$\mathcal{C}_{\ell}$ observed in our single realization of the full
sky was defined as
\begin{equation}
\mathcal{C}_\ell \equiv \frac{1}{2\ell+1} \sum_m \left|a_{\ell m}\right|^2 \label{eq:sky-cell-longhand}\textrm{,}
\end{equation}
and the theoretical power spectrum $C_{\ell}$ was taken to be the
expectation value of Eq. (\ref{eq:sky-cell-longhand}), $C_{\ell}= \langle
\mathcal{C}_{\ell} \rangle$.  We will henceforth use a shorthand for
such expressions, writing
\begin{align}
\mathcal{C}_\ell & \equiv \frac{a^{\dagger} \mathbf{\Delta}^\ell a}{2\ell + 1} \label{eq:sky-cell-shorthand} \textrm{;} \\ 
C_{\ell} & \equiv \frac{\Tr \mathbf{C\Delta}^\ell}{2\ell+1} \label{eq:cell-from-cov} \textrm{,}
\end{align}
where $\mathbf{C}=\langle a a^{\dagger} \rangle$ is the theory
covariance matrix and the elements of the $\mathbf{\Delta}^\ell$
matrices are
\begin{equation}
 \left(\mathbf{\Delta}^\ell\right)_{\ell' m', \ell'' m''}  = \delta^{\ell}_{\ell'}\delta^{\ell}_{\ell''} \delta_{m'm''}  \textrm{ (no sum).} \label{eq:define-delta-l}
\end{equation}
Thus $\mathbf{\Delta}^{\ell}$ is the projection operator into the
spin-$\ell$ subspace. The following two properties of
$\mathbf{\Delta}^\ell$ are useful:
\begin{align}
\Tr \mathbf{\Delta}^\ell & = (2\ell+1)\textrm{,} \\
\mathbf{\Delta}^\ell \mathbf{\Delta}^{\ell'} & = \delta_{\ell\ell'} \mathbf{\Delta}^{\ell} \textrm{ (no sum).}
\end{align}
Introducing the set of matrices $\mathbf{\Delta}^\ell$ produces
considerably more compact and readable equations at later stages.

\subsection{Expectation and variances}\label{sec:expect-vari}

Given the cut sky power spectrum estimates $\hat{C}_{\ell}$ defined by
equation (\ref{eq:c-ell-est}), we have respectively
\begin{align}
\langle \hat{C}_{\ell} \rangle & = \Tr \tilde{\mathbf{C}} \mathbf{R}^{\ell} \label{eq:estimator-bias}\\
V_{\ell\ell'} & \equiv \langle \hat{C}_{\ell}\hat{C}_{\ell'} \rangle - \langle \hat{C}_{\ell}\rangle \langle\hat{C}_{\ell'}\rangle = 2 \,\Tr \tilde{\mathbf{C}}\mathbf{R}^\ell \tilde{\mathbf{C}} \mathbf{R^{\ell'}} \label{eq:estimator-variance}
\end{align}
for the expectation and variance, where $\tilde{\mathbf{C}} =\langle
\tilde{a} \tilde{a}^{\dagger} \rangle= \mathbf{KCK}$ is the cut-sky
harmonic covariance matrix.

The estimator variance $V_{\ell\ell'}$ characterizes the random error
associated with estimating the ensemble quantity $C_\ell$ from a
single masked realization. This is the appropriate quantity for most
results in the paper and appendices. However, occasionally one wants a
measure of the extent to which the cut sky estimators accurately
predict the full sky (rather than ensemble averaged) power. A suitable
quantification is given by the following, which might be termed the
`cut-induced variance' (since it is necessarily zero on the full sky):
\begin{align}
\mathrm{CIV}_{\ell\ell'} & \equiv \left\langle \left(\hat{C}_\ell-  \mathcal{C}_\ell  \right)  \left( \hat{C}_{\ell'} -  \mathcal{C}_{\ell'}\right)\right\rangle =  2\,\Tr \mathbf{C} \mathbf{Z}^{\ell} \mathbf{C} \mathbf{Z}^{\ell'} \label{eq:civ} \\
\textrm{where } \mathbf{Z}^{\ell} & =  \tilde{\mathbf{R}}_{\ell} - \frac{\mathbf{\Delta}^\ell}{2\ell+1}\textrm{.}
\end{align}
The diagonal part of the cut-induced variance for a $20\%$ azimuthal
sky cut is plotted as a band in Figure \ref{fig:c-ratios}. It may be
verified that $\mathrm{CIV}_{\ell\ell'} \ne V_{\ell\ell'}$; expanding
expression (\ref{eq:civ}) shows that the cut-induced variance is equal
to the sum of the cut-sky and full-sky cosmic variances minus a unique
cross-term.

\subsection{The reconstruction matrices}

Let us now turn to specific reconstruction methods. The
$\mathbf{R}^\ell$ matrices for the PCL case
({\it e.g.} Ref. \cite{2001PhRvD..64h3003W}) read
\begin{align}
\mathbf{R}^\ell_{\mathrm{PCL}} & = \sum_{\ell'} (M^{-1})^{\mathrm{PCL}}_{\ell\ell'}\mathbf{\Delta}^{\ell'}/(2\ell'+1)\textrm{,} \label{eq:pcl-kernel} \\
\textrm{with } M^{\mathrm{PCL}}_{\ell\ell'} & = \Tr \mathbf{\Delta}^{\ell} \tilde{\mathbf{\Delta}}^{\ell'}/(2\ell'+1) \label{eq:pcl-debiaser}  \textrm{,}   
\end{align}
where the $2\ell'+1$ normalization on each of these expressions
is conventional.  It may readily be verified that these form unbiased
estimates for the full-sky, ensemble-averaged $C_{\ell}$'s in an
exactly isotropic theory since the covariance matrix $\mathbf{C}$ may
be written as
\begin{equation}
\mathbf{C} = \sum_{\ell} C_\ell \mathbf{\Delta}^\ell \label{eq:concordance-cov}\textrm{.}
\end{equation}
In this isotropic case one can write the covariance matrix
on the cut sky
\begin{equation}
\tilde{\mathbf{C}} = \sum_\ell C_\ell \tilde{\mathbf{\Delta}}^l \textrm{,}
\end{equation}
showing that our harmonic-space $\tilde{\mathbf{\Delta}}^\ell$ plays
the role of $\mathbf{P}^\ell$ in the notation of Tegmark's pixel-space
exposition of the QML estimator \cite{1997PhRvD..55.5895T}.
Accordingly, the QML reconstruction matrices are written\footnote{In
  equation (\ref{eq:qml-kernel}) and below we adopt the convention of
  assuming the existence of an inverse for the singular matrix
  $\tilde{\mathbf{C}}$.  Practically speaking one can regularize the
  matrix using an additive numerical noise term, or simply use the
  pseudo-inverse, since $\tilde{\mathbf{C}}^{-1}$ always appears
  conjugated by $\mathbf{K}$, the null directions of which lead to the
  uninvertibility.}:
\begin{align}
\mathbf{R}^\ell_{\mathrm{QML}} & = \sum_{\ell'} \left(M^{-1}\right)_{\ell\ell'} \tilde{\mathbf{C}}^{-1} \tilde{\mathbf{\Delta}}^{\ell'} \tilde{\mathbf{C}}^{-1} \label{eq:qml-kernel} \\
\textrm{where } M^{\mathrm{QML}}_{\ell\ell'} & = \Tr \tilde{\mathbf{C}}^{-1} \tilde{\mathbf{\Delta}}^\ell \tilde{\mathbf{C}}^{-1} \tilde{\mathbf{\Delta}}^{\ell'} \textrm{.} \label{eq:qml-deconvolve}
\end{align}
In these expressions it is possible to substitute for $\mathbf{C}$ a
false covariance matrix $\mathbf{C}^{\mathrm{conc}}$ which differs
from the true theory matrix used in expressions
(\ref{eq:estimator-bias}), (\ref{eq:estimator-variance}) and
(\ref{eq:civ}).  This represents the state of affairs when an
incorrect assumption is made by an analyst about the isotropy (or some
other aspect) of the underlying theory, as simulated in
Sec. \ref{sec:theories} above and Appendix \ref{sec:relat-betw-qml}
below. In numerical construction of the QML estimators we assumed a
variance on the monopole and dipole of $1000 \mu K^2$. This
effectively projects out information which is contaminated by
cross-talk from the monopole and dipole, and is likely to be
over-cautious, but residual foregrounds make it hard to quantify the
uncertainty in the WMAP zeroing of these quantities.  (See also the
discussion in Ref. \cite{1997PhRvD..55.5895T}.) We verified the
results were not sensitive to the precise variance assumed on
$\ell=0,1$.

As for the PCL case, the QML estimates are unbiased ($\langle
\hat{C}_{\ell}^{\mathrm{QML}} \rangle = C_{\ell}$) if both
$\mathbf{C}$ and $\mathbf{C}^{\mathrm{conc}}$ are isotropic. If
$\mathbf{C}=\mathbf{C}^{\mathrm{conc}}$ they are also optimal in the
sense that no unbiased estimator (quadratic or otherwise) can start
from the cut-sky $\tilde{a}_{\ell m}$'s and produce $C_{\ell}$
estimates with a smaller covariance ellipsoid
\cite{1997PhRvD..55.5895T}. 

However the QML estimator has sometimes been criticized for the
dependence of its optimality on the assumed covariance matrix -- it
appears to rely on the structure of the assumed underlying theory in a
way that the PCL estimator does not. [No $\mathbf{C}$ matrices appear
in expressions (\ref{eq:pcl-kernel}) and (\ref{eq:pcl-debiaser}).]
When the covariance matrix assumed may be incorrect, is it safer to
use the PCL estimator?  The answer is `no'; in fact the
anisotropy-induced errors in QML estimates are typically smaller than
those in PCL estimates.  We now explain why this should be the case.

\subsection{The relationship between QML~and~PCL~estimators}\label{sec:relat-betw-qml}

\begin{figure}
\begin{center}
\includegraphics[width=0.5\textwidth]{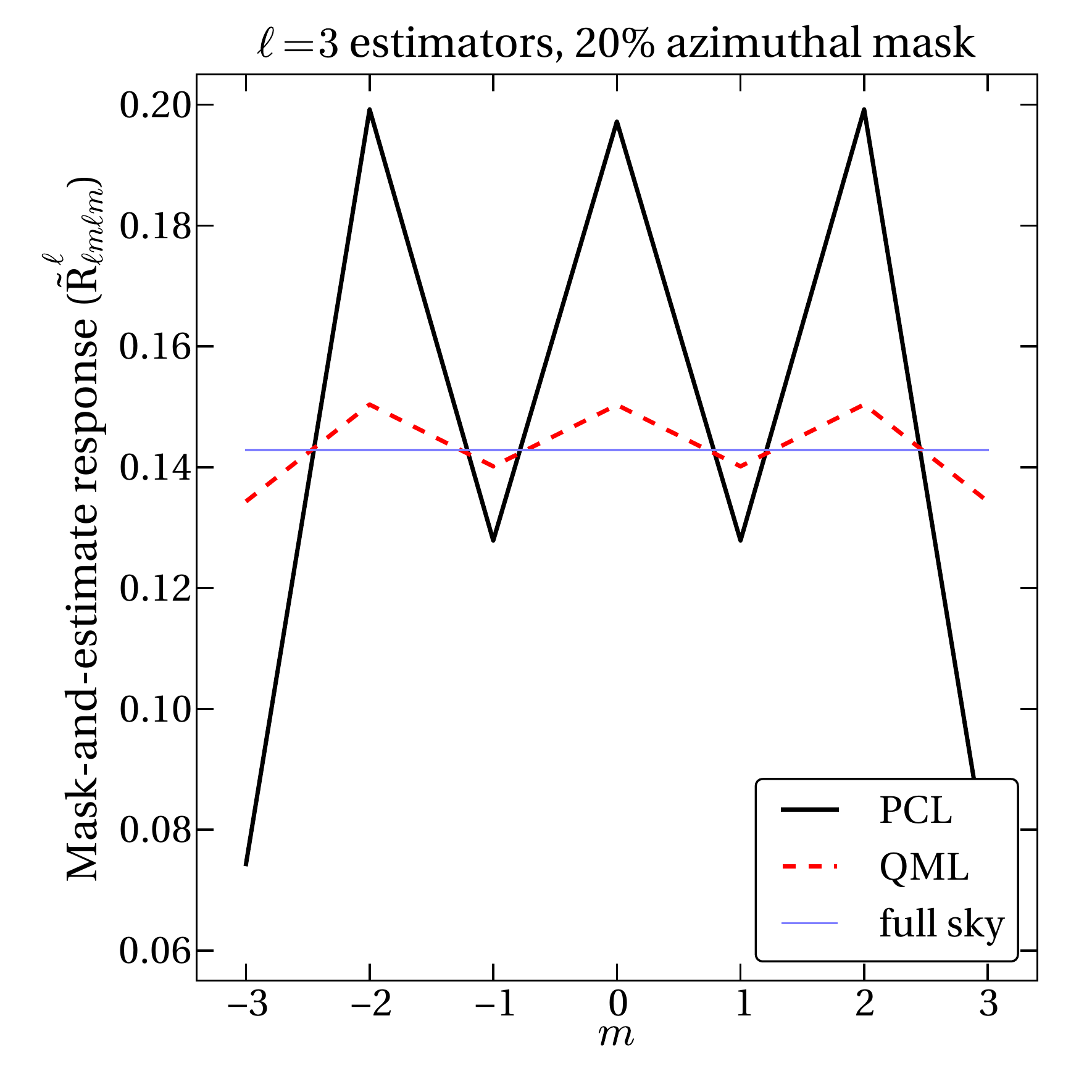}
\end{center}
\caption{(Color online.) The weighting which different estimators give
  to full-sky modes of differing $m$ in estimating the $C_{\ell}$'s
  (here illustrated for $\ell=3$ with a 20\% azimuthal mask). The full
  sky estimator (thin solid line) by definition weights each $m$
  equally; see Eq. (\ref{eq:sky-cell-longhand}).  Masking the sky
  then using the QML estimator (dashed line) comes close to
  reproducing this weighting, despite the loss of information
  associated with the first operation. The action of masking the sky
  then estimating $\ell=3$ power using the PCL technique (thick solid
  line) favours $|m|=2$ and $m=0$ while downweighting $|m|=3$ modes, 
  thus giving a less reliable estimate of the full sky
  power. Similar trends are seen at other $\ell$.
}\label{fig:m-response-l3}
\end{figure}

By examining the relationship between QML and PCL estimators, it is
possible to show that QML estimators (derived on the assumption of
isotropy) are statistically superior even if the underlying theory
breaks isotropy in an unknown way.  In outline, the QML estimator can
be understood as minimizing the cross-talk from variance in
neighbouring $\ell$ modes. It can only do this by having prior
information about the shape of the spectrum in the region of the
$\ell$ estimate under construction. But since the observed sky --
regardless of its isotropy -- has a $\mathcal{C}_{\ell}$ power
spectrum with a very similar shape to the theoretical model, the QML
estimator is expected to be superior to the PCL estimator under any
model compatible with the observed sky.

This argument does rely on the QML estimator not placing undue weight
at given $\ell$ on any particular $m$-mode. The full-sky estimator by
definition gives even weight to each $m$ [see expression
(\ref{eq:sky-cell-longhand})]. To accurately reproduce power spectra,
cut-sky estimators must trade off equal weighting of the $m$ modes
against down-weighting modes which are particularly contaminated by
mask mode-coupling.  Derived on the assumption of an isotropic theory,
it is not clear whether the QML estimator will do a better or worse
job than the PCL estimator in this limited sense. A calculation shows,
however, that the QML is superior -- it actually weights the full-sky
$m$-modes more evenly than the PCL estimator.  This is illustrated in
Figure \ref{fig:m-response-l3}, which shows the weight given to each
$m$ mode on the full sky under the composite operation of
masking-then-estimating. The weights are flatter for the QML estimator
(dashed line) than for the PCL estimator (thick solid line). It is
actually a fortuitous result of the shape of the concordance power
spectrum that this is true; otherwise the reliability of the QML
estimator would depend more sensitively on the underlying anisotropic
theory.

We now demonstrate the crucial result that the PCL and QML estimators
become identical for a flat power spectrum. The covariance matrix is
then proportional to the identity, $\mathbf{C} = \alpha \mathbf{I}$,
so that the QML reconstruction matrices reduce to

\begin{align}
\mathbf{R}^{\ell}_{\mathrm{QML}} & = \alpha^{-2}\,
\mathbf{M}_{\ell\ell'}^{-1} \tilde{\mathbf{\Delta}}^{\ell'} = \alpha^{-2}\,
\mathbf{M}_{\ell\ell'}^{-1} \mathbf{K\Delta^{\ell'}K} \label{eq:QML-becomes-PCL}\\
\mathbf{M}_{\ell\ell'}^{\mathrm{QML}} &= \alpha^{-2}\, \Tr \tilde{\mathbf{\Delta}}^\ell \tilde{\mathbf{\Delta}}^{\ell'}  = \alpha^{-2}\, \Tr \mathbf{\Delta}^\ell \tilde{\mathbf{\Delta}}^{\ell'} \label{eq:quasi-PCL-M}
\end{align} 
where the final expression for $\mathbf{M}^{\mathrm{QML}}$ is obtained
by expanding the masked expressions ($\tilde{\mathbf{\Delta}}^{\ell} =
\mathbf{K\Delta^\ell K}$) and using the condition
$\mathbf{K}^2=\mathbf{K}$ obtained in Section \ref{sec:cut-sky}.

By rewriting equation (\ref{eq:c-ell-est}) as $\hat{C}_{\ell} =
\vec{a}^{\dagger} \mathbf{KR^\ell K} \vec{a}$, it follows that the
$\mathbf{R}^\ell$ appearing in any statistical expression must arise
in the combination $\mathbf{K R^\ell K}$. This means that one can,
without loss of generality, dispense with the explicit masking
$\mathbf{K}$ matrices in expression (\ref{eq:QML-becomes-PCL}), again
relying on the identity $\mathbf{K}^2 = \mathbf{K}$. Finally, the
$\alpha^2$ factors in equations (\ref{eq:QML-becomes-PCL}) and
(\ref{eq:quasi-PCL-M}) may be mutually cancelled, since $\mathbf{M}$
appears only in the expression for $\mathbf{R}$. Thus we may write
\begin{align}
\mathbf{R}^\ell_{\mathrm{QML}} & \sim \mathbf{M}_{\ell\ell'}^{-1} \mathbf{\Delta}^{\ell'} = \mathbf{R}^\ell_{\mathrm{PCL}} \\
\textrm{where } \mathbf{M}_{\ell\ell'} & = \Tr \mathbf{\Delta}^\ell \tilde{\mathbf{\Delta}}^{\ell'}
\end{align}
where the $\sim$ symbol should be read as `yields identical estimates
to' -- {\it i.e.} it denotes an equivalence relation, not an approximate
equality of the matrix elements. To verify this, compare the above
with Eqs (\ref{eq:pcl-kernel},~\ref{eq:pcl-debiaser}), noting that the
missing factors of $2\ell'+1$ are conventional normalizations which
exactly cancel between the two lines.

This demonstrates that, if $\mathbf{C}= \alpha \mathbf{I}$, QML
estimates are identical to PCL estimates. The result does not rely
on any assumptions about the Galactic cut being small. However, for a
small Galactic cut, the mask operation $\mathbf{K}$ acquires a narrow
banded structure at high-$\ell$ such that each $\ell$ is effectively
coupled only to a finite range of $\ell'$ from $\ell-\Delta \ell$ to
$\ell+\Delta \ell$. Thus, even though the concordance covariance
matrix is not proportional to the identity, at high $\ell$ its
relevant, local structure can be adequately approximated as such.
This demonstrates the equivalence of the QML and PCL estimators in
this regime.

To understand the difference between QML and PCL estimators one can
reverse the argument above [{\it i.e.} one replaces
$\mathbf{\Delta}^{\ell'}$ by $\tilde{\mathbf{\Delta}}^{\ell'}$ in
equation (\ref{eq:pcl-kernel}) then compares with equation
(\ref{eq:qml-kernel}), finding the latter simply pre-weights the
data].

\begin{figure}
\begin{center}
\includegraphics[width=0.5\textwidth]{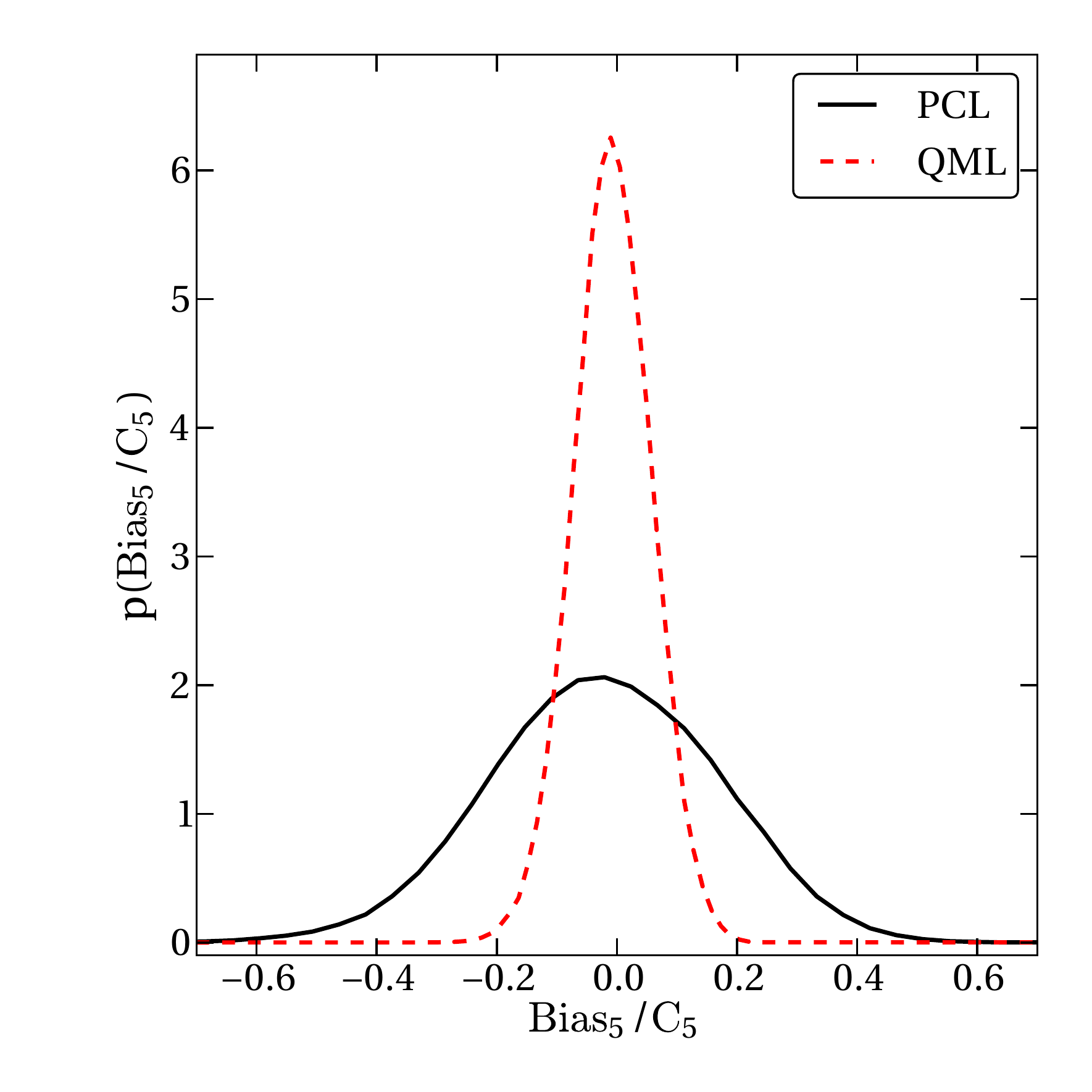}
\end{center}
\caption{(Color online.) An example of how the QML estimator typically
  remains superior to the PCL estimator is given by comparing, for
  $200\,000$ random anisotropic theories, the bias on the cut-sky
  power spectrum estimates (here illustrated for $\ell=5$ with a 20\%
  azimuthal mask). The width of the curves show that, for a given
  theory, the QML estimator (dashed curve) is typically significantly
  less biased than the PCL estimator (solid curve). Similar results
  hold at other $\ell$'s and for the extra variance induced by the
  anisotropy. }\label{fig:sim-bias}
\end{figure}

It follows from all this that the QML estimator can, roughly speaking,
be expected to remain superior to the PCL estimator for any theory
compatible with our sky. To demonstrate this explicitly, we draw
random covariance matrices with power spectrum  equal to that
of the observed ILC, but taking a random distribution of power between
different $m$ modes. Explicit calculations for each of these theories
show that the QML biases (\ref{eq:estimator-bias}) and variances
(\ref{eq:estimator-variance}) are significantly smaller than their PCL
counterparts. A specific illustration is given in Figure
\ref{fig:sim-bias}, where we plot a histogram of the biases on
$\hat{C}_5^{\mathrm{PCL}}$ (solid line) and $\hat{C}_5^{\mathrm{QML}}$
(dashed line) for $200\,000$ random theories. The PCL estimator has a
significantly broader distribution of biases than the QML estimator,
showing that the QML technique typically produces more reliable
estimates of the full sky power.

In conclusion, for any anisotropic theory which is compatible with our
observed sky, estimates for the power spectrum formed using the QML
technique (despite being derived assuming isotropy) are expected to be
superior to PCL estimates for the same quantity.

\section{Estimators for $C(\theta)$ and $S_{1/2}$}\label{sec:estim-ctheta-s_12}

In this section we demonstrate that the pixel-based cut-sky 
correlation function,
\begin{equation}
\mathcal{C}(\theta)^{\mathrm{cut}} \equiv \frac{\int \dd \n1 \dd \n2 M(\n1) M(\n2) T(\n1) T(\n2)\, \delta_\theta}{\int \dd \n1 \, \dd \n2 M(\n1) M(\n2)\, \delta_\theta}\textrm{,}\label{eq:c-cut-appendix}
\end{equation}
where $\delta_\theta = \delta(\n1\cdot\n2 - \cos \theta)$, is
identical to the PCL-based estimator\footnote{We are grateful to
  A. Challinor for initially drawing our attention to this
  equivalence.}
\begin{equation}
\mathcal{C}(\theta)^{\mathrm{PCL}} \equiv \frac{1}{4\pi} \sum_\ell (2\ell+1) \hat{C}_\ell^{\mathrm{PCL}} P_\ell(\cos \theta) \textrm{.} \label{eq:c-cut-pcl-appendix}
\end{equation}
This has been stated before \cite{2004PhRvD..69h3524A} and is implicit
in other works \cite{2000astro.ph.10256S,2004MNRAS.350..914C}, but an
explicit demonstration has not, to our knowledge, appeared in the
literature. The argument holds for any weighting function
$M(\hat{\vec{n}})$: if $M$ takes values other than $0$ and $1$ the
harmonic-space matrix $\mathbf{K}$ [still defined by equation
(\ref{eq:mask-matrix-def})] is no longer idempotent, but no part of
the proof below is affected by such a change.

Equation (\ref{eq:c-cut-appendix}) may be expressed
\begin{align}
\mathcal{C}(\theta)^{\mathrm{cut}} & = \frac{1}{F(\theta)} \sum_\ell \frac{2\ell+1}{4 \pi} \tilde{C}_{\ell} P_{\ell}(\cos \theta)\textrm{,}
\end{align}
where $F(\theta)$, equal to the denominator of
(\ref{eq:c-cut-appendix}), is a normalizing function dependent only on
$M(\hat{\vec{n}})$. Here $\tilde{\mathcal{C}}_{\ell}$ is the power
spectrum of the masked (or weighted) sky,
\begin{equation}
\tilde{C}_{\ell} = \frac{\tilde{\vec{a}}^{\dagger} \mathbf{\Delta}^{\ell} \tilde{\vec{a}}}{2\ell + 1}= M_{\ell\ell'}^{\mathrm{PCL}} \hat{C}_{\ell'}^{\mathrm{PCL}}\textrm{,}
\end{equation}
where $\mathbf{M}^{\mathrm{PCL}}$ is defined by equation
(\ref{eq:pcl-debiaser}). A power spectrum can be calculated from
$\mathcal{C}(\theta)^{\mathrm{cut}}$:
\begin{align}
\hat{C}_{\ell}^{\mathrm{cut}} & = 2\pi \int_{-1}^1 \mathcal{C}(\theta)^{\mathrm{cut}} P_{\ell} (\cos \theta) \, \dd \cos \theta \\ 
& = T_{\ell\ell'} \tilde{C}_{\ell'} = T_{\ell\ell'} M_{\ell'\bar{\ell}}^{\mathrm{PCL}} \hat{C}_{\bar{\ell}}^{\mathrm{PCL}}  \textrm{,} \label{eq:cl-cut-cl-pcl-relation}
\end{align}
where the matrix
\begin{equation}
T_{\ell\ell'} = \frac{2\ell'+1}{2} \int_{-1}^1 \frac{P_\ell(\cos \theta) \, P_{\ell'}(\cos \theta)}{F(\theta)} \dd \cos \theta \label{eq:T-matrix}
\end{equation}
depends only on $F(\theta)$ (and hence only on the sky cut, not any
aspects of the theory or realization).

If we temporarily consider a theory which is isotropic, we have
$\langle \mathcal{C}(\theta)^{\mathrm{cut}} \rangle = C(\theta) =
\langle \mathcal{C}(\theta)^{\mathrm{PCL}} \rangle$ and hence, by
linearity, $\langle \hat{C}^{\mathrm{cut}}_{\ell} \rangle = \langle
\hat{C}^{\mathrm{PCL}}_{\ell}\rangle$.  But then, comparing with
expression (\ref{eq:cl-cut-cl-pcl-relation}), the only possibility is
that the matrix $T_{\ell\ell'}$ is the inverse of the matrix
$M_{\ell\ell'}^{\mathrm{PCL}}$ -- in other words that
\begin{equation}
\hat{C}^{\mathrm{cut}}_{\ell} = \hat{C}^{\mathrm{PCL}}_{\ell}\textrm{.}
\end{equation}
We reiterate that neither $\mathbf{T}$ nor $\mathbf{M}^{\mathrm{PCL}}$ depend on
either the underlying theory nor the particular realization in hand,
and therefore this result is independent of isotropy. Finally, one
inverts the Legendre transform to gain the desired result,
\begin{equation}
\mathcal{C}(\theta)^{\mathrm{PCL}} \equiv \mathcal{C}(\theta)^{\mathrm{cut}}\textrm{,}
\end{equation}
valid for any theory. It follows immediately that $\shalfcut$ derived
from expression (\ref{eq:shalf-cut-pcl}) must be mathematically
equivalent to pixel-based estimates.

The above proof depends on the invertibility of
$M^{\mathrm{PCL}}_{\ell\ell'}$. It is well known that this matrix is
not invertible for all choices of sky-cut (although for all masks
considered in the present work we have found it to be
well-behaved). However, in any limit where $\det
\mathbf{M}^{\mathrm{PCL}} \to 0$, one must have $\det \mathbf{T} \to
\infty$.  According to definition (\ref{eq:T-matrix}), this will occur
if and only if $F(\theta) \to 0$ for some $\theta$ -- in other words
if and only if the cut sky contains, in the limit, no two points
separated by certain values of $\theta$. It follows that, whenever the
entire correlation function can be recovered from the cut sky, the PCL
estimates can be made and the relationship proved above holds.

\subsection*{Aside: $\shalfcut$ is biased high}

We should note in passing that, because $\shalf$ is
quadratic in the $\mathcal{C}_\ell$'s, its expectation value does not
follow simply by replacing the $\mathcal{C}_\ell$'s with the
$C_{\ell}$'s in Eq. (\ref{eq:shalf-quadratic}); rather, the full sky
expectation value reads for the concordance theory
\begin{equation}
S_{1/2} \equiv \langle \shalf \rangle = \sum_{\ell \ell'} s_{\ell \ell'} \left( C_\ell C_\ell' + \frac{2 C_\ell^2 \delta_{\ell \ell'}}{(2\ell+1)} \right) \label{eq:shalf-theory} \textrm{.}
\end{equation}
The second term in Eq. (\ref{eq:shalf-theory}) contributes very
significantly to the expectation value, which breaks down term-by-term
as $S_{1/2} = (4.9 + 3.7) \times 10^4 = 8.6 \times 10^4 \, \mu
\mathrm{K}^4$ for the WMAP5 best fit $C_{\ell}$'s.  This means that
(for instance) the comparison of our full sky with the theory values
in Table~1 of Ref. \cite{2008arXiv0808.3767C} is not strictly
appropriate; with the cosmic variance term included, the observed
$\shalf$ statistic is made to look even more discrepant with the
theory.  The expectation value of $\shalfcut$ calculated directly from
the cut sky $\hat{C}_\ell$ is also increased, for similar reasons, but
by a larger amount corresponding to the larger variance on the cut sky
power spectrum estimates:
\begin{equation}
\langle \hat{C}_{\ell} \hat{C}_{\ell'}\rangle =
\langle \mathcal{C}_{\ell} \mathcal{C}_{\ell'}\rangle - \frac{2 C_{\ell}^2 \delta_{\ell \ell'}}{(2\ell+1)} + V_{\ell\ell'} \textrm{,}
\end{equation}
where $V_{\ell\ell'}$ is the variance of the $\hat{C}_{\ell}$'s, given
by expression (\ref{eq:estimator-variance}).  This biases cut-sky
$\shalf$ values to be higher than their full-sky
counterparts, for instance by $\sim 8300\, \mu \mathrm{K}^4$ for PCL
and $\sim 1100\, \mu \mathrm{K}^4$ for QML reconstructions with a
$20\%$ azimuthal sky cut.

At face value, such biases make it more surprising that the measured
$\shalf$ should be so small and $\shalfcut$ even smaller. However, the
standard deviation of $\shalfcut$ is very large ($\sim 10^5\, \mu
\mathrm{K}^4$) so that the biases do not have a significant impact on
the frequentist significances. Furthermore, and regardless of the
magnitude of the biasing, the Monte Carlo techniques used in Refs
\cite{2008arXiv0808.3767C} and \cite{2009arXiv0911.5399E} are anyway
valid (they automatically take the biasing into account). We have
therefore included the discussion above only for pedagogical interest.

\section{Designer theory}\label{sec:designer-theories}

In Section \ref{sec:designer-theory-main} we used a theory with
covariance matrix $\mathbf{C}$ determined by two considerations:
\begin{enumerate}
\item The full sky power spectrum is given by $C_{\ell} =
  \mathcal{C}_{\ell}^{\mathrm{ILC}}$, where
  $\mathcal{C}_{\ell}^{\mathrm{ILC}}$ is the observed power spectrum
  on the full-sky ILC map;
\item The cut sky power spectrum (PCL estimator using a $20\%$
  azimuthal cut) is biased, {\it i.e.} its expectation value does not equal
  the full sky power; instead we set $\langle
  \hat{C}^{\mathrm{PCL}}_{\ell} \rangle = C_{\ell}^{\mathrm{cut}}$. To
  reproduce the causes of our own sky's low $\shalfcut$, we set
  $C_{\ell}^{\mathrm{cut}}=0$ for $\ell=2, 3, 5$ and $7$; at all other
  $\ell$, $C_{\ell}^{\mathrm{cut}} =
  \mathcal{C}_{\ell}^{\mathrm{ILC}}$.
\end{enumerate}

Both constraints are linear in the full sky covariance matrix [see
Eqs. (\ref{eq:cell-from-cov}) and (\ref{eq:estimator-bias})].  While
one can construct a matrix $\mathbf{C}$ satisfying these constraints
using straight-forward linear algebra, the result is not unique and
furthermore it is hard to enforce that $\mathbf{C}$ be positive
definite (as it must be to define a valid covariance matrix).
Therefore we adopted the package {\sf CVXOPT}\footnote{{\tt
    http://abel.ee.ucla.edu/cvxopt/}; this package performs convex
  optimization within a cone. (The space of positive definite matrices
  is an example of a cone in this sense.)} to find a suitable theory
$\mathbf{C}$ within the set of positive definite matrices.  {\sf
  CVXOPT} allows us to find a unique solution by minimizing any convex
quadratic form, for which we chose the function $\Tr
\mathbf{C}^2$. The choice at first appears arbitrary; but
schematically, by considering the eigenvalues of $\mathbf{C}$, one can
imagine that minimizing $\Tr \mathbf{C}^2$ tries to `equalize power
between as many modes as possible'. This in turn is motivated by our
attempt to minimize the cosmic variance on $\shalfcut$, leading to the
most peaked likelihood function (and hence best possible likelihood
gains over the concordance theory).

These statements can be made somewhat more mathematically concrete,
but we did not find a full proof that minimizing $\Tr \mathbf{C}^2$
minimizes the cosmic variance of $\shalfcut$. Instead, Section
\ref{sec:designer-theory-main} gave a strict lower bound on the
variance of $\shalfcut$, and stated that our theory comes close to
saturating this limit. The remainder of the present appendix explains
the origin of such a variance floor.

We start by considering, for simplicity, the full sky $\shalf$.  We
also temporarily approximate the $\mathcal{C}_{\ell}$ likelihood
function as Gaussian. Both of these simplifications will be removed in
due course; in particular, all of our numerical results use the exact
likelihood. The variance of the $\shalf$ statistic may be written
\begin{equation}
\langle (\shalf)^2 \rangle - \langle \shalf \rangle^2 = 4  {\vec{c}}^{\top}  \mathbf{sV^{\mathrm{fs}}s}  {\vec{c}} + 2\, \Tr \mathbf{sV^{\mathrm{fs}}sV^{\mathrm{fs}}}\textrm{,}\label{eq:shalfvar-gaussian}
\end{equation}
where $\vec{c}$ is a vector composed of the $C_{\ell}$'s [as
defined by equation (\ref{eq:cell-from-cov})], and $\mathbf{V}^{\mathrm{fs}}$
represents the full sky cosmic variance,
\begin{equation}
V^{\mathrm{fs}}_{\ell\ell'} = \frac{2 \, \Tr \mathbf{C\Delta^{\ell} C\Delta^{\ell'}}}{(2\ell+1)(2\ell'+1)} \textrm{.}
\end{equation}
We wish to minimize Eq. (\ref{eq:shalfvar-gaussian}) with respect to
$\mathbf{C}$ while keeping $C_{\ell}$ constant. Using a standard
Lagrange multipler technique, one obtains 
\begin{align}
 \sum_{\ell_1 \ell_4} s_{\ell_1 \ell_2} s_{\ell_3 \ell_4} (C_{\ell_1} C_{\ell_4} + V^{\mathrm{fs}}_{\ell_1 \ell_4}) \mathbf{\Delta^{\rm \ell_2} C \Delta^{\rm \ell_3}} & =0\textrm{,} \label{eq:off-diag-minimiz} 
\end{align}
where $\ell_2 \ne \ell_3$; and
\begin{align}
  C_{\ell} \mathbf{\Delta}^{\ell} & = \mathbf{\Delta^{\ell} C \Delta^{\ell}} \label{eq:on-diag-minimiz}\textrm{.}
\end{align}
The most obvious solution to the minimization equations
(\ref{eq:off-diag-minimiz}) and (\ref{eq:on-diag-minimiz}) is the
isotropic one,
\begin{equation} 
\mathbf{C} = \sum_{\ell} C_{\ell} \mathbf{\Delta}^{\ell}\textrm{.}\label{eq:isotropic-min-var}
\end{equation}
One can verify that the solution (\ref{eq:isotropic-min-var}) is a
minimum (not maximum) of expression (\ref{eq:shalfvar-gaussian}).  To
demonstrate that no other minima exist, consider the only alternative
to (\ref{eq:isotropic-min-var}): namely that $\mathbf{\Delta}^{\ell_2}
\mathbf{C \Delta}^{\ell_3} \ne 0$ and equation
(\ref{eq:off-diag-minimiz}) is instead satisfied by making the
numerical coefficient vanish. We consider the case where this is true
for all $\ell_2$, $\ell_3$ ($\ell_2 \ne \ell_3$), but the ideas
generalize straight-forwardly to the case with only limited numbers of
non-zero off-diagonal terms. The most general solution is
\begin{equation}
V_{\ell\ell'} = -C_{\ell} C_{\ell'} + \sum_i \lambda_i Q^i_{\ell \ell'} \textrm{,} \label{eq:violating-variance}
\end{equation}
where the $\mathbf{Q}^i$ are symmetric matrices which satisfy
$\mathbf{s}\mathbf{Q}^i \mathbf{s}=0$ (off-diagonal) and
\begin{equation}
\sum_i \lambda_i Q^i_{\ell \ell} = \frac{2\ell+3}{2\ell+1} C_{\ell}^2 \label{eq:violating-variance-constraint}\textrm{.}
\end{equation}
The $\mathbf{Q}^i$ may be found numerically using a singular value
decomposition technique.

Let us consider whether a physical (positive definite) solution to
equations (\ref{eq:violating-variance}) and
(\ref{eq:violating-variance-constraint}) exists. A necessary condition
is that
\begin{equation}
  |V_{\ell\ell'}| \le \frac{2 C_{\ell}
  C_{\ell'}}{\sqrt{(2\ell+1)(2\ell'+1)}}\textrm{.}
\end{equation}
This condition is violated by equation (\ref{eq:violating-variance})
with $\lambda_i=0$, but can a suitable choice of $\lambda_i$ remove
the violation?  There are far fewer $\mathbf{Q}^{i}$ matrices than
degrees of freedom in $\mathbf{V}^{\mathrm{fs}}$, so that one would
need a numerical coincidence to be able to remove the violation
simultaneously at all $\ell$. We verified computationally that, for
our choice of $C_{\ell}$, this is indeed not possible.

Now when the Gaussian simplification is abandoned,
expression (\ref{eq:shalfvar-gaussian}) picks up extra terms of the form
\begin{align}
s_{\ell_1\ell_2} s_{\ell_3\ell_4} \Tr \mathbf{C} \mathbf{\Delta}^{\ell_1} \Tr
\mathbf{C\Delta^{\ell_2} C\Delta^{\ell_3} C\Delta^{\ell_4}}  \textrm{ and} \\
s_{\ell_1\ell_2} s_{\ell_3\ell_4}  \Tr
\mathbf{C \Delta^{\ell_1} C\Delta^{\ell_2} C\Delta^{\ell_3} C\Delta^{\ell_4}}\textrm{.}
\end{align}
One may verify that with these terms, equation
(\ref{eq:isotropic-min-var}) remains a local minimum of the $\shalf$
variance. Since there are no other valid local minima in the Gaussian
approximation, we should not expect to find new local minima appearing
in the non-Gaussian case (the simplification modifies only the
skewness, not the width, of the likelihood).  A full analytic
calculation is prohibitive, but the numerical results quoted in the
main paper use the full, non-Gaussian likelihood.

Now let us consider the cut sky case. The de-biased PCL estimates $\hat{\mathbf{c}}$
are related to the power spectrum of the cut sky $\tilde{\mathbf{c}}$ by
\begin{equation}
\hat{\mathbf{c}} = \mathbf{M}^{-1} \tilde{\mathbf{c}}
\end{equation}
for the matrix $\mathbf{M}$ defined by equation
(\ref{eq:pcl-debiaser}). (The `PCL' superscript is dropped for
concision.) Because of the exact relation
\begin{align}
\shalfcut & = \sum_{\ell\ell'} \hat{C}^{\mathrm{PCL}}_{\ell} \hat{C}^{\mathrm{PCL}}_{\ell'} s_{\ell\ell'}  \nonumber \\
 & = \sum_{\ell_1 \ell_2 \ell_3 \ell_4} M^{-1}_{\ell_1 \ell_2} M^{-1}_{\ell_3 \ell_4} \tilde{\mathcal{C}}_{\ell_2} \tilde{\mathcal{C}}_{\ell_4} s_{\ell_1 \ell_3}  \label{eq:shalf-cut-lengthy}
\end{align}
(see Appendix \ref{sec:estim-ctheta-s_12}) we may define
\begin{equation}
\tilde{\mathbf{s}} = (\mathbf{M}^{-1})^{\top} \mathbf{s} \mathbf{M}^{-1}\textrm{,}
\end{equation}
so that equation (\ref{eq:shalf-cut-lengthy}) simplifies to
\begin{equation}
\shalfcut = \sum_{\ell \ell'} \tilde{s}_{\ell \ell'} \tilde{\mathcal{C}}_{\ell} \tilde{\mathcal{C}}_{\ell'}\textrm{.}\label{eq:shalf-cut-from-cutsky}
\end{equation}
The cut-sky reasoning then follows through exactly as for the full-sky
case, except with $\mathbf{s}$ and $\mathbf{C}$ replaced by
$\tilde{\mathbf{s}}$ and $\tilde{\mathbf{C}}$ respectively. For the
lower bound theory one obtains
\begin{equation}
\tilde{\mathbf{C}} = \sum_{\ell\ell'} M_{\ell\ell'} C_{\ell'}^{\mathrm{cut}} \mathbf{\Delta}^{\ell}\textrm{.} \label{eq:unobtainable-cutsky-cov}
\end{equation}
Clearly this ignores the implicit restrictions on $\tilde{\mathbf{C}}$
arising from its status as a cut-sky, rather than full-sky, covariance
matrix. (Specifically, a valid $\tilde{\mathbf{C}}$ must live in the
cut-sky subspace so that $\tilde{\mathbf{C}} = \mathbf{K\tilde{C}K}$.)
However the set of all valid $\mathbf{\tilde{C}}$ is, crucially, a
subset of the positive-definite matrices which were considered for the
full-sky case.  Therefore our test theory still gives a lower bound
for the set of valid theories.

To actually calculate the lower bound we draw 20\,000 sets of $\tilde{a}_{\ell
  m}$'s according to the covariance matrix
(\ref{eq:unobtainable-cutsky-cov}) and, for each, calculate the power
spectrum $\tilde{\mathcal{C}}_{\ell}$ and hence $\shalfcut$ according
to equation (\ref{eq:shalf-cut-from-cutsky}).  Calculating the
variance on this random sample leads to the numerical lower limit
quoted in Section \ref{sec:designer-theory-main}.

Finally note that, because $\mathbf{M}$ is almost diagonal, our
Monte Carlo results are almost equivalent to those obtained by
calculating $\shalf$ in an isotropic theory satisfying
\begin{equation}
\mathbf{C} = \sum_{\ell} C_{\ell}^{\mathrm{cut}} \mathbf{\Delta}^{\ell}\textrm{.}
\end{equation}
This is the justification for our intuitive explanation that the lower
bound on the variance of $\shalfcut$ is given by the variance of
$\shalf$ in an isotropic, full-sky theory with power spectrum equal to
the cut-sky power spectrum of the designer theory.

\section{A numerical problem and solution}\label{sec:harmonic-space-masking}

There is a hidden numerical pitfall in the harmonic-space masking
operation as defined by Eq. (\ref{eq:mask-matrix-def}). The matrix
$\mathbf{K}$ is not band-limited, which means that truncating at
finite $\ell$ produces cut-sky vectors $\tilde{\vec{a}}$ which retain
some information about the data inside the cut. This garbled
information is visible in maps as low-amplitude ringing around the
edges of the cut. The QML estimator, in particular, is very efficient
at regenerating the full sky from this trace of unwanted information.

We investigated two methods of mitigating this problem, both of which
generated results in good agreement with pixel space techniques. The
first is heuristic, simply smoothing the input (full sky) and output
(cut sky) maps to angular scales larger than
$180^{\circ}/\ell_{\mathrm{max}}$.

The second, which we adopted for our final results\footnote{This
  method was suggested to us by S. Gratton.}, is to use an eigenvector
decomposition. For a specified sky fraction $f$, we calculate
$\mathbf{K}$ to finite $\ell_{\mathrm{max}}$ and find its eigenvalues
and vectors. We then replace the smallest eigenvalues (specifically, a
fraction $f$ of the eigenvalues) by zero and all other eigenvalues by
one. The final operation is then guaranteed to be idempotent (unlike
the ad hoc smoothing approach) and also discards exactly the right
fraction of information from the input map.  Visually, we found the
maps produced looked almost identical to those masked in pixel
space. As commented above, we verified that the final estimator
results produced from a harmonic-space analysis were closely
compatible with those produced from a pixel space analysis. The latter
are slow and cumbersome [they cannot take advantage of simplification
(\ref{eq:azimuthal-simplification})], but do not suffer from the
band-limitation problem and therefore serve as a useful point of
comparison.

\section{Rapid calculation of $L^2_{\mathrm{max}}$}\label{sec:rapid-calc-l2_m}

In the main text, we discussed the planarity of $\ell=2$ and $\ell=3$
power in the observed CMB. This is uncovered
\cite{deOliveiraCosta:2003pu} by computing the `angular momentum dispersion'
\begin{equation}
L_\ell^2 = \frac{\sum_m m^2 \left|a_{\ell m}\right|^2}{\ell^2 \sum_m \left|a_{\ell m}\right|^2} \textrm{,}\label{eq:std-l2-definition}
\end{equation}
designed to detect `planarity' of power. By maximizing this quantity
across different rotations of the sky, one produces a preferred
direction in which the power is most planar. 
Frequentist anomalies are
then reported if the sky-measured maximum values of $L_\ell^2$ have
small $P$-values according to Monte Carlo simulations of statistically
isotropic skies -- or if the maximizing directions for two different
$\ell$'s are coincident.

We will show below that calculating $L_\ell^2$ after rotating the sky
by Euler angles\footnote{We adopt the convention of
  \cite{1960quthanmo.book.....V}; an Euler rotation $(\alpha, \beta,
  \gamma)$ successively rotates the physical sky relative to the
  fixed, right-handed coordinate system by $-\alpha$, $-\beta$ and
  $-\gamma$ around the $z$, $x$ and $z$ axes respectively. The final
  $z$ rotation would not affect the value of $L_\ell^2$, so we fix
  $\gamma=0$.} ($\phi-\pi/2$,$-\theta$,0) is equivalent to forming the
quantity
\begin{equation}
L_\ell^2(\theta,\phi) = \frac{ n_i n_j \sum_{mm'} L^{mm'}_{ij} a_{\ell m} a_{\ell m'}^*}{\ell^2 \sum_m \left|a_{\ell m}\right|^2} \label{eq:l2-algebraic}
\end{equation}
where the vector $\vec{n}=\left(\sin \theta \cos \phi, \sin \theta
  \sin\phi, \cos \theta\right)$ and the matrix elements $L^{mm'}_{ij}$
are given explicitly below. Hence, for a given set of $a_{\ell m}$'s, the
problem of maximizing $L_\ell^2$ reduces to finding the maximal
eigenvector of the $3 \times 3$ real symmetric matrix $\sum_{mm'}
L^{mm'}_{ij} a_{\ell m} a_{\ell m'}^*$. This algorithm is cheaper by orders of
magnitude than numerical maximization methods that appear to have been
used to date.

We now prove relation (\ref{eq:l2-algebraic}) and give explicit forms
for the matrix elements $L_{ij}^{mm'}$. Consider the spin-$\ell$
function $\Psi_\ell$, which may be expanded as
\begin{equation}
\Psi_\ell(\theta,\phi) = \sum_m a_{\ell m} Y_{\ell m}(\theta,\phi) \textrm{.}
\end{equation}
By considering the angular momentum of this function in the $z$
direction,
\begin{equation}
(\vec{e}_z \cdot \hat{\vec{J}})\Psi_\ell(\theta,\phi) = -i \partial_{\phi} \Psi_\ell(\theta,\phi) = \sum_m m a_{\ell m} Y_{\ell m}(\theta,\phi) \textrm{,}
\end{equation}
where $\hat{\vec{J}}$ is the fiducial angular momentum operator and
$\vec{e}_{z}$ is the unit vector in the $z$ direction, one may rewrite
equation (\ref{eq:std-l2-definition}) as
\begin{equation}
L_\ell^2 = \frac{\left\langle \Psi_\ell \, \left| \, \left( \vec{e}_z \cdot \hat{\vec{J}} \right)^2  \right| \, \Psi_\ell \right\rangle}{\ell^2 \langle \Psi_\ell | \Psi_\ell \rangle} \textrm{,} \label{eq:almost-there}
\end{equation}
 where the inner product is defined as usual:
\begin{equation}
\langle \Phi | \Psi \rangle = \int \Phi^*(\theta,\phi) \Psi(\theta,\phi)\,\sin \theta\, \dd \theta\, \dd \phi \textrm{.} 
\end{equation}

To convert expression (\ref{eq:almost-there}) for a single $L_\ell^2$ value
into the form (\ref{eq:l2-algebraic}), which is claimed to give all possible sky-rotated values, 
we first need to see that the following two operations are equivalent:

\begin{enumerate}

\item rotating the function by angle $\alpha$ about an axis specified
  by the vector $\vec{r}$, then measuring $L_\ell^2$;
\item rotating the $z$ coordinate direction by angle $-\alpha$ about $\vec{r}$
  while keeping the function fixed, {\it i.e.} replacing $\vec{e}_z$ in equation (\ref{eq:almost-there})
  by the rotated vector $\vec{e}_z^{\phantom{z}\prime}$.
\end{enumerate}
The equivalence is intuitively clear because the only preferred direction in expression (\ref{eq:almost-there})
is given by the $\vec{e}_z$ vector. To establish the result more formally, one can use the
commutation relations for the angular momentum operators applied to an infinitesimal rotation, and then 
extend, as usual, to finite rotations by exponentiation.

% One can proceed as follows. Consider the effect of an infinitesimal rotation of the vector $\Psi_\ell$:
% %
% \begin{eqnarray}
% | \Psi_\ell \rangle & \to & \left(1 + i \epsilon \vec{r} \cdot \hat{\vec{J}}\right) | \Psi_\ell \rangle \textrm{,} \\
% \langle \Psi_\ell | & \to & \langle \Psi_\ell | \left(1-i\epsilon \vec{r} \cdot \hat{\vec{J}}\right) \textrm{,}
% \end{eqnarray}
% %
% where $\epsilon$ is a small angle and $\vec{r}$ gives the axis of rotation. Then using
% the standard commutation relations for $\vec{J}$ gives
% %
% \begin{eqnarray}
%  \vec{e}_z \cdot \hat{\vec{J}}\, | \Psi_\ell \rangle & \to & (1+i\epsilon \vec{r} \cdot \hat{\vec{J}})\left(\vec{e}_z-\epsilon \vec{e}_z \times \vec{r}\right) \cdot \hat{\vec{J}}\, | \Psi_\ell \rangle \textrm{,}\hspace{1cm} \\
% \langle \Psi_\ell | \vec{e_z} \cdot \hat{\vec{J}} & \to & \langle \Psi_\ell | (\vec{e}_z - \epsilon \vec{e}_z \times \vec{r} ) \cdot \hat{\vec{J}} \left(1-i\epsilon \vec{r} \cdot \hat{\vec{J}} \right) \textrm{.}
% \end{eqnarray}
% %
% The $L^2_\ell$ value of the rotated
% function is then
% %
% \begin{equation}
% L_\ell^2 = \frac{\left\langle \Psi_\ell \, \left| \, \left( \left[\vec{e}_z - \epsilon \vec{e}_z \times \vec{r} \right] \cdot \hat{\vec{J}}\right)^2 \, \right| \, \Psi_\ell \right\rangle}{\langle \Psi_\ell | \Psi_\ell \rangle} \textrm{.}
% \end{equation}
% which is equivalent to the second operation defined above.

The result is that
\begin{equation}
L_\ell^2(\theta,\phi) = \frac{\left\langle \Psi_\ell \left| \left( \vec{n} \cdot \hat{\vec{J}}\, \right)^2 \right| \Psi_\ell \right\rangle}{\ell^2 \langle \Psi_\ell | \Psi_\ell \rangle} = \frac{n_i n_j \left \langle \Psi_\ell \left| \hat{J}_i \hat{J}_j \right| \Psi_\ell \right \rangle}{\ell^2 \langle \Psi_\ell | \Psi_\ell \rangle}\textrm{,}
\end{equation}
with $\vec{n}$ defined as above. In harmonic space the operator $\hat{J}_i \hat{J}_j$ forms the matrix
elements appearing in Eq.~(\ref{eq:l2-algebraic}):
\begin{equation}
L_{ij}^{mm'} = \langle Y_{\ell m'} | \hat{J}_i \hat{J}_j | Y_{\ell m} \rangle = \sum_{m''}  J^{mm''}_i J^{m''m'}_j \textrm{,}
\end{equation}
where the harmonic space angular momentum operators are obtained
numerically using the relations
%by conjugating $\hat{J}_z$ with an
%appropriate rotation operator:
\begin{align}
J^{mm'}_z  & =  m' \delta_{mm'}\textrm{;} \nonumber \\
J^{mm'}_x &= \frac{1}{2} \left(J_+^{mm'} + J_-^{mm'}\right)\textrm{ and} \nonumber \\
J^{mm'}_y &= \frac{1}{2i} \left(J_+^{mm'} - J_-^{mm'}\right)\textrm{.}
%\hat{J}^{mm'}_x & =  \sum_{m''} m'' \, D_\ell^{mm''}(0,\tfrac{-\pi}{2},\tfrac{\pi}{2},) \, D_\ell^{m''m'}(\tfrac{-\pi}{2},\tfrac{\pi}{2},0)\textrm{;} \hspace{1cm} \nonumber \\
%\hat{J}^{mm'}_y & =  \sum_{m''} m'' \, D_\ell^{mm''}(0,\tfrac{\pi}{2},0) \, D_\ell^{m''m'}(0,\tfrac{-\pi}{2},0) \textrm{.}
\end{align}
Here the matrix elements of the ladder operators $\hat{J}_{\pm}$ obey
\begin{equation}
J_{\pm}^{mm'} = \sqrt{l(l+1)-m(m+1)} \delta_{m\pm 1,m'}\textrm{.}
\end{equation}

Finally we note that analytic moments of the statistical distributions may be
calculated using the above formalism; however, in practice, the distributions
are rather asymmetrical at low $\ell$ and it is conceptually and computationally
easier to use Monte Carlo results -- this is extremely fast using the new
algorithm, especially since the $L_{ij}^{mm'}$ matrix elements can be
precomputed and cached for each $\ell$ of interest.

\end{document}